\documentclass[10p]{article}
\usepackage{amssymb}
\usepackage[english]{babel}
\usepackage[colorlinks=false]{hyperref}
\usepackage{amsthm}
\usepackage{easyReview}
\usepackage{float} 
\usepackage{booktabs}
\usepackage{amsmath}
\usepackage{tabularx}
\usepackage{booktabs}
\usepackage{pifont}
\usepackage{graphicx}
\usepackage{authblk}

\usepackage{geometry}
\title{Multilayer public transport networks}

\author[1]{Tina Šfiligoj$^*$} %% Author name

\affil[1]{Faculty of Maritime Studies and Transport, University of Ljubljana, Slovenia, e-mail: tina.sfiligoj@fpp.uni-lj.si}

\author[2]{Renzo Massobrio} %% Author name

\author[3]{Oded Cats}

\affil[2]{Department of Electronics-ICT, Faculty of Applied Engineering, University of Antwerp, Belgium}

\affil[3]{Department of Transport \& Planning, Delft University of Technology, The Netherlands}

\begin{document}

\maketitle

% \begin{center}
%      {\Large Multilayer public transport networks} 
%      % Multilayer public transport networks: from network taxonomy to model applications
% \end{center}
% \vspace{0.35cm}
% \begin{large}
%     \begin{center}
%         Tina Šfiligoj, Renzo Massobrio, Oded Cats
%     \end{center}
%     \end{large}

% \vspace{2cm}

\subsection*{Abstract}

The introduction of network science approaches into public transport research has seen great advances in the past 15 years. However, it has become apparent that monolayer networks are often not sufficient to model and analyse real-world systems in sufficient detail. In the last decade, the theory of multilayer networks has proven to be an invaluable tool in various disciplines, including transport. Multilayer networks consist of layers of networks that are coupled among themselves. This enables modelling of complex systems with heterogeneous elements and relations between them.  
Although there is a body of work in public transport research that uses multilayer networks, the related literature is scattered, lacking unified terminology and agreed-upon approaches. We posit that there is vast uncovered potential in using multilayer network approaches to public transport modelling, planning, and operations. We first present the basic formalisms of multilayer networks with a focus on how they (may) relate to public transport networks. We then provide a systematic review of the literature on multilayer networks in public transport research. We identify and taxonomise ways in which public transport systems are modelled as multilayer networks. Based on the survey and drawing from the state and history of network science in public transport research as well as multilayer approaches across other application domains, we propose a research agenda for multilayer public transport networks for the upcoming decade(s).

\medskip

\textbf{Keywords:} public transport, network science, multilayer networks, transport planning

% \subsection*{Highlights}

% \begin{itemize}
%     \item A brief introduction to the theory and methods of multilayer networks (MLNs).  

%     \item Survey of the existing literature on multilayer public transport networks (MLPTNs).  

%     \item A taxonomy of MLPTNs based on the guiding principles of layering the PTNs.  

%     \item A research agenda ranging from methodological to PTN planning problems. 
% \end{itemize}

% \tableofcontents

 \section{Introduction}

 \subsection{The bigger picture}

The scientific area of complex networks is a relatively young science, having emerged with the works of Watts and Strogatz in 1998 \cite{watts1998collective}	and Barabási and Albert in 1999 \cite{barabasi1999emergence}. Despite its young age, it has seen tremendous advances in the understanding of complex socio-physical systems. 
Among those are urban systems, and as they grow in scale and complexity, the ability to describe them as interconnected networks of elements has become increasingly important.
The timeliness of complex network approaches to modern problems has led to statements of complexity science being the science of the 21st century, the role of which was attributed to physics in the 20th century.

As crucial as the introduction of the representation of systems as networks was, it has become apparent in recent years that representations limited to one type of nodes and edges, which constitute the basis of network science, often do not suffice to faithfully reflect the underlying complexity of the considered systems. This has led to advanced network models, of which multilayer networks (MLN) are among the most prominent. In MLNs, a system is modelled as a set of interconnected ordinary networks, each constituting a separate layer, where nodes and edges in each layer may represent different entities and/or connections between them. The discipline of multilayer networks is an emerging area in both theory and applications. The seminal review articles on multilayer networks by Kivelä et al. \cite{kivela2014multilayer} and Boccaletti et al. \cite{boccaletti2014structure} were both published in 2014 and presented the first attempts to systematise research into various versions of interconnected networks. Since then, the area of multilayer networks has burst and has seen significant theoretical and applied advances in the last decade \cite{aleta2019multilayer, aleta2025multilayer}. Among the most prominent application domains for MLNs are biology \cite{haas2017designing, finn2019use, gosak2018network}, neuroscience \cite{de2017multilayer, vaiana2020multilayer}, power networks \cite{kazim2025analysis}, and transport and communication networks \cite{crainic2022taxonomy, li2025review, alessandretti2023multimodal}.

% \cite{silk2023conceptual}, \cite{pilosof2017multilayer}, 

Multilayer networks also play an increasingly important role in transport research, where pioneering research was on air transport network analysis \cite{du2016analysis}. Other areas of transport where MLNs are adopted are maritime transport \cite{alderson2020analysis}, inter-regional freight and passenger transport \cite{li2025review}, and public transport \cite{gallotti2014anatomy}.

\subsection{Public transport networks}

Public transport networks (PTN) are an epitome of complex systems, interconnecting the infrastructure and operational characteristics of the underlying transport systems. This is evident from a large body of literature, where they are taken as real-world examples of complex networks. Together with their role as a backbone of sustainable urban mobility, the need for transport researchers to connect the two areas is evident. 

\begin{figure}[H]
	\begin{center}
		\texttt{}\includegraphics[scale=0.4]{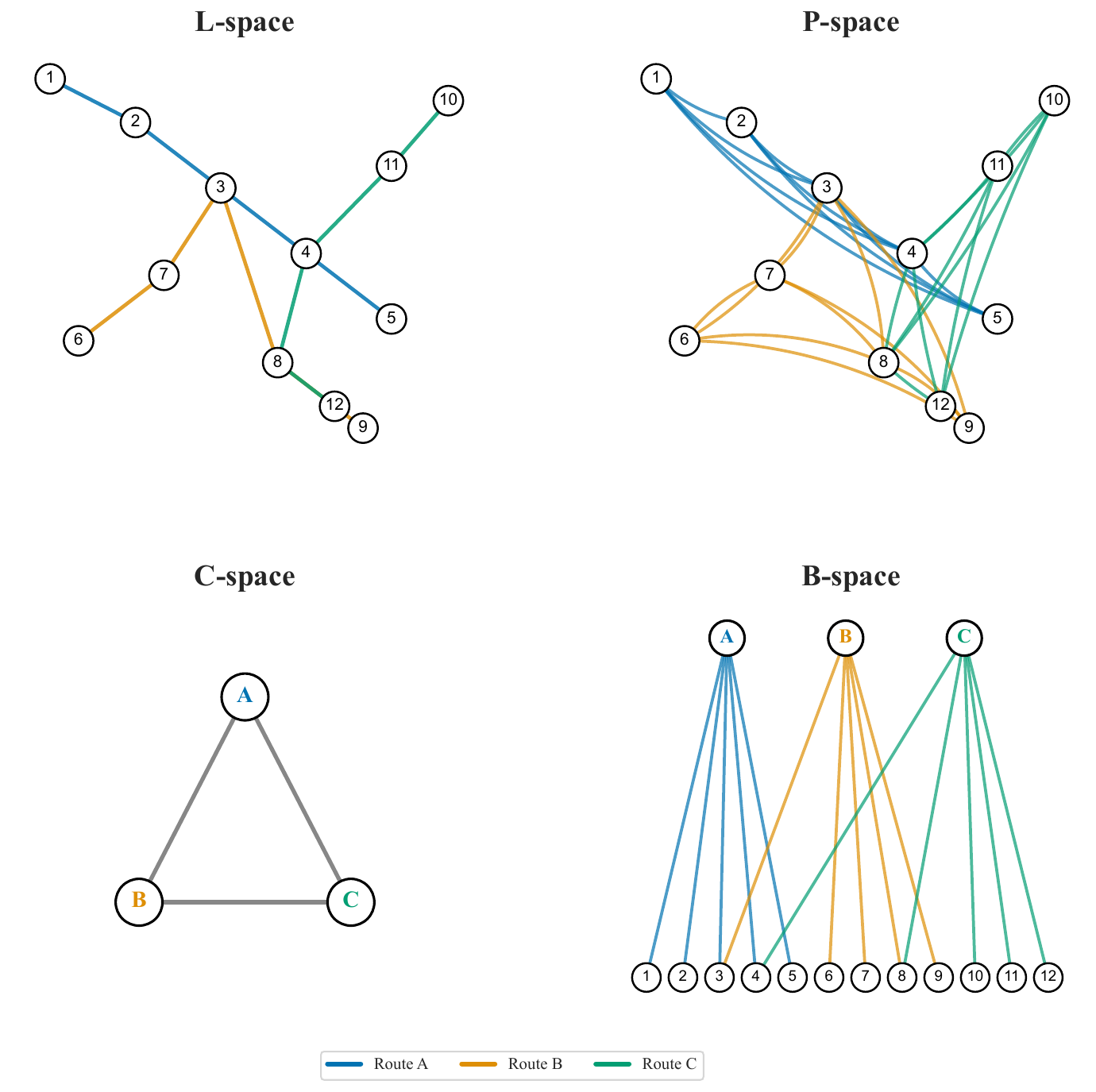}
	\end{center}
	\caption{An illustration of the standard PTN graph representations.}
	\label{fig:greps}
\end{figure}

Network science methods were first introduced into public transport (PT) research in a structured manner by von Ferber et al. in 2009 in their seminal work \cite{von2009public}. Building on the previous work of Sienkiewicz et al. \cite{sienkiewicz2005statistical} who in 2005 analysed structural properties of 22 Polish cities, they proposed the by now-standard graph representations of PTNs (L-, P-, C- and B-space) and analysed their topological properties, focusing on small-world phenomena in public transport. The representations are briefly described below and illustrated in Figure \ref{fig:greps}.

\begin{itemize}
    \item \textbf{L-space}, or space-of-infrastructure \cite{luo2020can}: two nodes are connected with an edge if they represent consecutive stops on a line. Weighted representations often include in-vehicle times as edge weights.
    \item \textbf{P-space}, or space-of-service \cite{luo2020can}: two nodes are connected if they are serviced by the same line (i.e., each line is a clique).  Weighted representations often include service frequency or waiting times as edge weights.
    \item \textbf{C-space}: nodes represent lines and there is an edge between two nodes if the corresponding lines share a stop.
    \item \textbf{B-space}: a bipartite representation where the two sets of nodes represent stops and lines, respectively. There is an edge between two nodes if the corresponding stop lies on the corresponding line.
\end{itemize}

In the following years, there was an increasing interest in connecting network science and PT, albeit primarily from the physicists' perspective, who used PTNs as real-world examples of complex networks. An early call to bridge this gap in favour of transport research was made by Derrible and Kennedy in 2011 \cite{derrible2011applications} where they proposed several ways towards more transport-oriented topics. Starting in the mid 2010s, transport researchers have increasingly employed network science methods. This has led to original insights into the properties of PTNs and gave rise to novel ways of PT planning and design, mainly from two focus areas: first, network structure in terms of its topological properties (e.g. \cite{dimitrov2016method, de2019public}), and second, PT resilience analysis (e.g.,  \cite{zhang2015assessing, cats2016robustness, chopra2016network, cats2020metropolitan, massobrio2024topological}).

We point out three notable observations pertaining to the integration of network science into (public) transport research. First, the adoption of network science approaches has been relatively slow compared to other application areas, such as biology or social networks. Second, the transport problems addressed are disproportionately focused on a small subset of relevant issues, such as network structure and resilience. Finally, although the use of network science methods in PT research has reached a level of maturity and the variety of its applications to a wide range of transport problems continues to increase, several crucial issues remain open. It has become increasingly evident that basic network approaches are insufficient to capture the full complexity of PTNs. We keep this in mind as we turn our focus to multilayer public transport networks.

 \subsection{Multilayer public transport networks}

One of the most promising ways to address the last observation is the adoption of multilayer networks in PT research and planning. An obvious, although by no means the only one, intersection of MLNs and PTNs are multi-modal PT systems where each mode is represented as a separate layer and the layers are interconnected to represent the coupling of different modes. This is confirmed in the existing literature, where those represent the most widely studied topic. Notwithstanding, the adoption of MLN approaches is lagging behind compared to research in other application domains. Moreover, related work commonly investigated the statistical properties of multilayer transport networks, much in the same way that research has progressed for monolayer networks. Robustness-related studies are by far the most prevalent, again drawing a parallel with the inclusion of (monolayer) network science into PT research, while neglecting other PT planning and operations problems. At the moment, applications in public transport are rare and scattered, with limited agreed-upon definitions or conventions.
We address these limitations by considering the following research questions.

\begin{enumerate}
    \item What are the existing approaches to multilayer networks in various application domains and how can we taxonomise those?
    \item What is the range of applications for which multilayer networks have been developed in public transport research?
    \item In what conceptually distinct ways can public transport networks be modelled as a multilayer network?
    \item What are the implications of multilayer network representations for public transport planning and operations?  
\end{enumerate}

We attempt to answer these research questions by providing the following:

\begin{itemize}
    \item A brief introduction to mathematical formalism in the modelling and analysis of multilayer networks.
    \item A structured and critical review of previous work on multilayer networks in public transport research. 
    \item Based on the previous point, a taxonomisation of multilayer public transport networks and a proposal for a standardised terminology.
    \item A research agenda for MLPTNs.
\end{itemize}

The remainder of this article is structured as follows. First, in Section \ref{sec:basics}, we introduce fundamental concepts, notation and mathematical formalisms from the theory of multilayer networks. Section \ref{sec:litrev} presents the results of the structured literature search. A classification of approaches is presented, with the main contribution being the taxonomisation of multilayer representations of PTNs in Section \ref{sec:tax}. Based on the theory and findings from the previous sections, Section \ref{sec:agenda} proposes a research agenda for multilayer public transport networks. Section \ref{sec:concl} offers closing remarks.

\section{Basics of multilayer networks}
\label{sec:basics}

Multilayer networks were first systematically considered in seminal articles by Kivelä et al. \cite{kivela2014multilayer} and Boccaletti et al. \cite{boccaletti2014structure}. Both studies review research up to 2014 and make an effort to unify the widely scattered terminology in the field. The research into multilayer networks has burst since their publication more than a decade ago.

In this section, we first present some basic terminology (\ref{sec:btn}). We then introduce a classification of MLNs (\ref{sec:mlntypes}). This is followed by mathematical formalisms (\ref{sec:mfmln}), and an inventory of some of the most common metrics for MLNs (\ref{sec:metrics}).

\subsection{Basic terminology and notations}
\label{sec:btn}

We first recall some basic definitions for monolayer networks. A graph $G(V,E)$ is a mathematical object composed of a set of vertices (nodes) $V=\{v_i\}$, $|V|=n$ and a set of edges $E=\{e_{ij}\}$, $e_{ij}=(v_i, v_j)$, $|E|=m$. $G$ has dimension (i.e., number of nodes) $n$ and size (i.e., number of edges) $m$. In an undirected graph, $e_{ij} = e_{ji}$ for all $i,j$, which is not generally the case in a directed graph. Edges can be assigned weights $w_{ij}\in \mathbb{R}$, representing either connection strength or cost. A graph is usually represented in matrix form by its adjacency matrix $A$ with elements $a_{ij}=1$ if there is an edge between nodes $v_i$ and $v_j$, and $a_{ij}=0$ otherwise. In the case of weighted networks, the weight matrix $W$ with elements $w_{ij}$ is used instead of $A$. Node degree is the number of its direct neighbours (equivalently, the number of edges incident to it) and can be calculated as $d_i = \sum_j a_{ij}$. Similarly, the node strength $s_i$ is obtained from the weight matrix as $s_i = \sum_j w_{ij}$.

Often, networks with one type of nodes and one type of edges are not sufficient for a realistic description of a complex system where distinct types of elements are interconnected and there exist several categories of relationships between them. This is where multilayer networks come in: multilayer networks are higher-level graph structures consisting of interconnected \textit{layers}, where each layer $\alpha$ is an ordinary network. Node $v_i$ in layer $\alpha$ is denoted by $v_i^\alpha$.
Each layer $\alpha$ contains a subset of nodes $N^{\alpha} \subseteq N$. The set of layers is denoted by $\mathcal{L}$, and $L = |\mathcal{L}|$ is the number of layers in the MLN.

\begin{figure}[H]
	\begin{center}
		\texttt{}\includegraphics[scale=0.65]{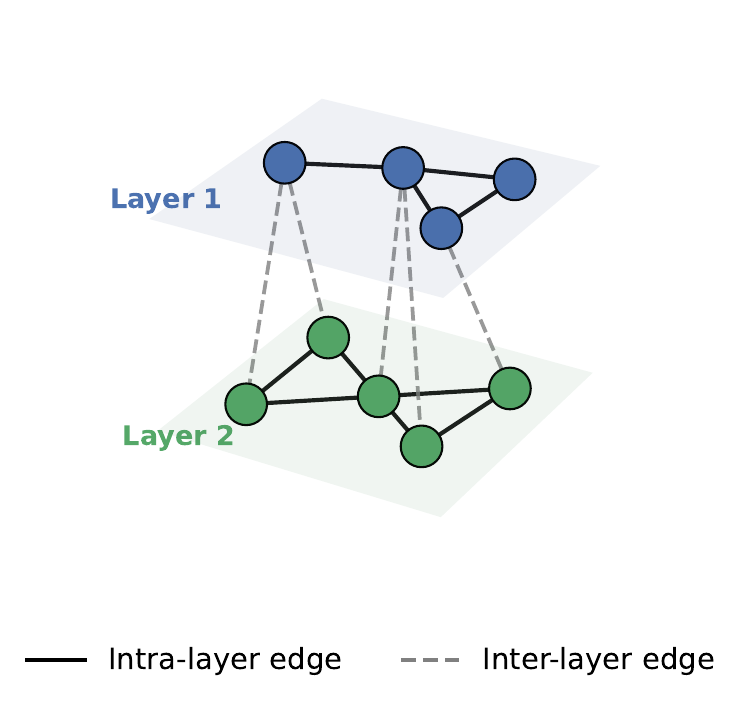}
	\end{center}
	\caption{An illustration of a simple multilayer network consisting of two layers. Intra-layer edges within a single layer are shown in bold lines and inter-layer edges connecting nodes from separate layers are shown in dashed lines.}
	\label{fig:mlntoy}
\end{figure}

In terms of edges, we distinguish between \textbf{intra-layer edges} within a layer, and \textbf{inter-layer edges} that connect nodes from separate layers. Intra-layer edges $e_{ij}^{\alpha} = (v_i^\alpha, v_j^\alpha)$ are those that are found in monolayer networks.
On the other hand, an inter-layer edge $e_{ij}^{\alpha\beta} = (v_i^\alpha, v_j^\beta), \alpha \ne \beta$ connects the node $v_i$ in layer $\alpha$ to the node $v_j$ in layer $\beta$. Inter-layer edges are also referred to as couplings between nodes in different layers. An illustration of a multilayer network is provided in Figure \ref{fig:mlntoy}.

\subsection{Types of multilayer networks}
\label{sec:mlntypes}

There are various ways to represent systems as multilayer networks and the representations will differ based on the nature of the system under consideration as well as aspects of the system that are of interest. The main point to stress is that there is no universally agreed-upon classification of MLNs. The original coarse-grained classification is that in \cite{kivela2014multilayer} into: $(i)$ edge-coloured MLNs, where nodes represent the same physical entities in all layers, and edges represent a different type of interaction in each layer, and $(ii)$ node-coloured MLNs where nodes (and edges) in different layers represent different entities (and/or relations between them). This division is at the core of any specific classification used by researchers in a specific area, but the level of classification is usually more detailed. Edge-coloured networks are mostly synonymous with multiplex networks, which are a prominent category of MLNs and are the most coherently described class of MLNs. Node-coloured networks in contrast exist under various names such as interconnected networks, interdependent networks, networks of networks, coupled networks, hypernetworks, multilevel networks, to name just a few. An additional complication is that different names are often used for the conceptually same structure, or the same name is used for different structures. Examples of different classifications are evident in key review articles in different domains (e.g., \cite{pilosof2017multilayer, hammoud2020multilayer, hamed2024comprehensive}).

\begin{figure}[H]
	\begin{center}
		\texttt{}\includegraphics[scale=0.325]{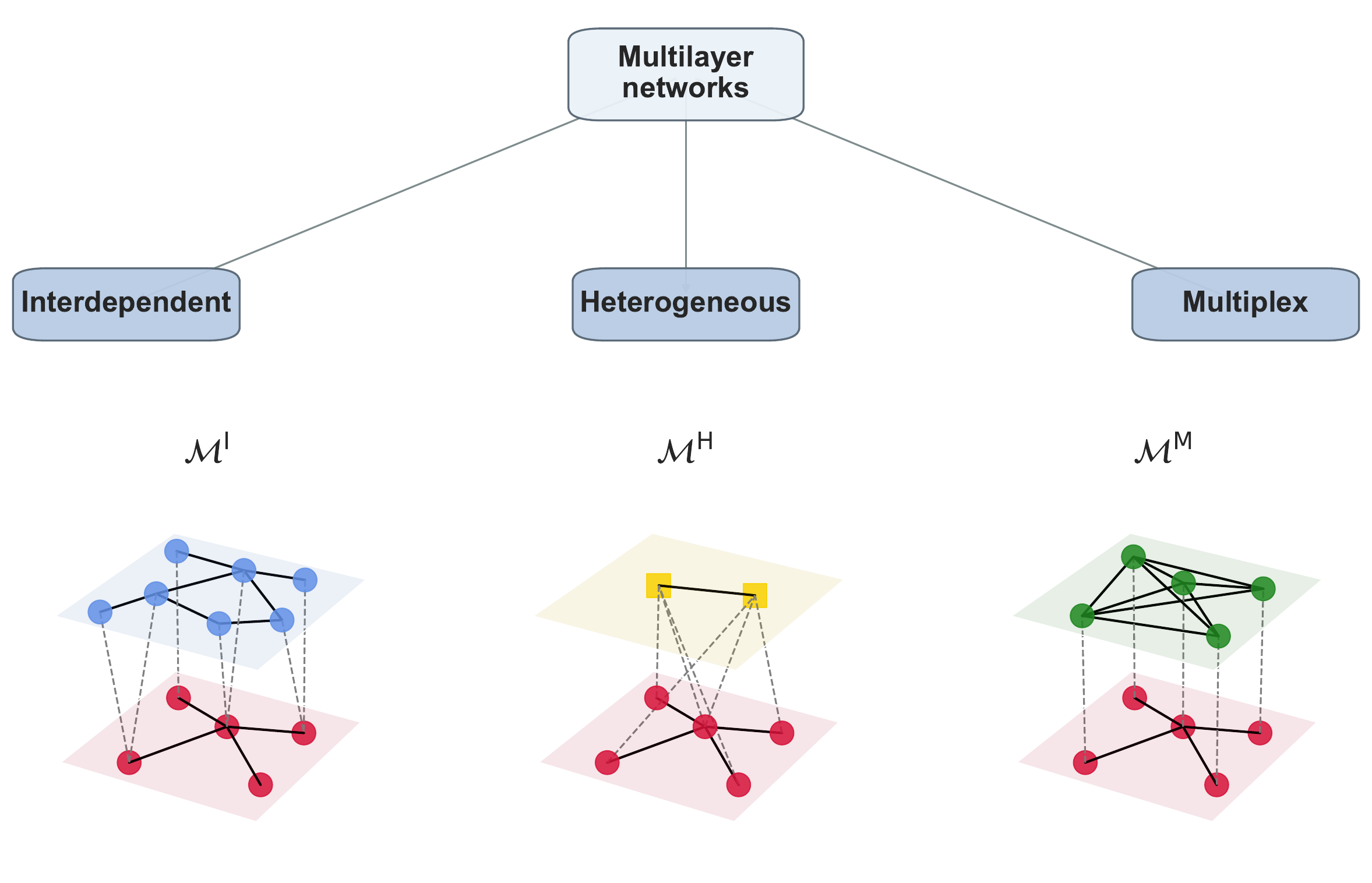}
	\end{center}
	\caption{Types of multilayer networks. The upper plot shows the classification tree of MLNs as presented in the this review. Below each category is an example of a multilayer public transport network. The leftmost plot represents $\mathcal{M}^{I}$ represents an interdependent network where the bottom layer is the L-space metro and top layer L-space bus. The middle network $\mathcal{M}^{H}$, is an example of a heterogeneous network, where the bottom layer is again the L-space metro and the top layer is the C-space representation of the same network.
    The rightmost network, represents a multiplex network $\mathcal{M}^{M}$ where the bottom layer is the L-space representation of the toy metro network and the top layer is the OD layer. (See also descriptions in the text).}
	\label{fig:mlntypes}
\end{figure}

In this work, we settle for a classification that generally conforms with the literature and decide upon a level of granularity that best reflects the needs from the PT perspective. We adopt the view that in general, multilayer networks can be divided into two main categories: multiplex networks and interconnected networks \cite{aleta2019multilayer}. We further distinguish between interdependent and heterogeneous networks as different categories of interconnected networks, as described in the following. For each of the three categories, we provide a PT example. The classification tree and the corresponding examples are shown in Figure \ref{fig:mlntypes}. For illustration of the presented concepts, we will consider a toy city with a public transport system as follows. The PT system consists of a metro network with 5 stations which are served by two lines, and a bus network with 7 stations served by 3 lines. We assume we have complete information on the system infrastructure and operation, as well as origin-destination (OD) passenger flows. 

\begin{itemize}
    \item \textbf{Interdependent networks}: Layers represent different systems with nodes representing different types of entities, but the systems are interconnected through some dependency.
    In interdependent networks, nodes are qualitatively similar. An example is a multi-modal PT system, where each layer represents a separate mode, and nodes in each layer represent the stations of the respective mode. In all cases, the stations serve the same function but differ in details (e.g., exact location, vehicle capacity, passenger capacity). Inter-layer edges can connect any pair of nodes and can naturally be assigned a weight (e.g., walking time from a bus to a metro station).

    The leftmost network $\mathcal{M}^{I}$ in Figure \ref{fig:mlntypes} is an example of an interdependent network. The bottom layer is a metro network and the top layer is an L-space representation of a bus network. The inter-layer edges connect nodes that represent stations and stops within walking distance.

    \item \textbf{Heterogeneous networks:} Nodes represent fundamentally different types of entities. Those include, but are not limited to, networks that represent different levels of aggregation, such as stations in one layer and lines in another layer in the case of PTNs. 
        
        An example in public transport systems is an extended B-space representation and is shown in the second network in  Figure \ref{fig:mlntypes}, $\mathcal{M}^{H}$: in the bottom layer the nodes represent stations, and in the top layer nodes represent routes. The inter-layer edges connect nodes that represent the stations lying on the respective routes.

    \item \textbf{Multiplex networks}: 
    The set of nodes N in a multiplex network represents a single set of physical entities. Each layer contains replicas of a subset of physical nodes, and represents a different type of interaction. Intra-layer edges represent the underlying type of connection between distinct nodes. 
    The nodes in different layers are connected with inter-layer edges if and only if they represent the same physical objects. In other words, there can exist inter-layer edges only between the replicas of the same physical node, $e_{ij}^{\alpha\beta}=1$ for $i=j$ and $\alpha\ne\beta$ and $e_{ij}^{\alpha\beta}=0$ otherwise. The couplings are categorical and represent the identity mapping between the replicas of the same physical node.

    An example of a multiplex PTN is the rightmost network in Figure \ref{fig:mlntypes}, $\mathcal{M}^{M}$. The bottom layer represents the L-space representation of a simple metro network, and the top layer represents the OD network. The nodes in both layers are replicas of the same physical nodes (i.e., metro stations), but each layer describes different relationships between them (i.e., infrastructural vs. passenger flows).
\end{itemize}

\subsection{Mathematical formulation of MLNs}
\label{sec:mfmln}

Whereas the structure of monolayer networks can be fully captured by connecting pairs of nodes by edges $e_{ij}$, which is translated into the form of the adjacency matrix, we see that in the case of multilayer networks with edges $e_{ij}^{\alpha\beta}$ we need four indices for their full description. This is naturally represented by \textit{tensors} of order 4 \cite{de2013mathematical}. A tensor is a higher-order generalisation of a matrix. An $m$-order tensor $\mathcal{A}$ has entries $a_{i_1i_2...i_m}$. A scalar (a number) is therefore a 0-dimensional tensor; a vector is a 1-dimensional tensor, and a matrix is a 2-dimensional tensor.

The element $a_{ij}^{\alpha\beta}=[\mathcal{A}]_{ij\alpha\beta}$ of the tensor is  $a_{ij}^{\alpha\beta}=1$ if there is an edge between node $i$ in layer $\alpha$ and node $j$ in layer $\beta$, and $a_{ij}^{\alpha\beta}=0$ otherwise. Similarly as in the case of monolayer networks, we can assign weights to edges. For example, in network $\mathcal{M}^{I}$ (Figure \ref{fig:mlntypes}), $a_{ij}^{11}=5$ would indicate that the in-vehicle time between the nodes $i$ and $j$ in the bottom (metro) layer is 5 minutes, and by $a_{ij}^{{12}}=3$ we would indicate that the walking time between the node $i$ in the bottom (metro) layer and the node $j$ in the upper (bus) layer is 3 minutes.

It should be noted that higher-order tensors are significantly richer structures than matrices.
Correspondingly, numerical methods on tensors exhibit prohibitive complexity, often exponential in the order of the tensor, and either polynomial or exponential in its size.
In order to address these challenges, several methods of multilayer networks simplification have been proposed. Tensor flattening to a supra-adjacency matrix 
% and layer aggregation are 
is among the most commonly used.
The supra-adjacency matrix $\mathbf{A}$ of a MLN is a block matrix, where diagonal blocks represent intra-layer edges (i.e. adjacency matrices of single layers $A^{\alpha_i}$) and off-diagonal blocks (i.e., coupling matrices $C^{\alpha_i\alpha_j}$) contain information on inter-layer couplings:

\begin{equation}
    \mathbf{A} = 
\begin{bmatrix}
A^{\alpha_1} & C^{\alpha_1\alpha_2} & \cdots & C^{\alpha_1\alpha_L} \\
C^{\alpha_2\alpha_1} & A^{\alpha_2} & \cdots & C^{\alpha_2\alpha_L} \\
\vdots & \vdots & \ddots & \vdots \\
C^{\alpha_L\alpha_1} & C^{\alpha_L\alpha_2} & \cdots & A^{\alpha_L} \\
\end{bmatrix}
\end{equation}

\begin{figure}[H]
	\begin{center}
		\texttt{}\includegraphics[scale=0.35]{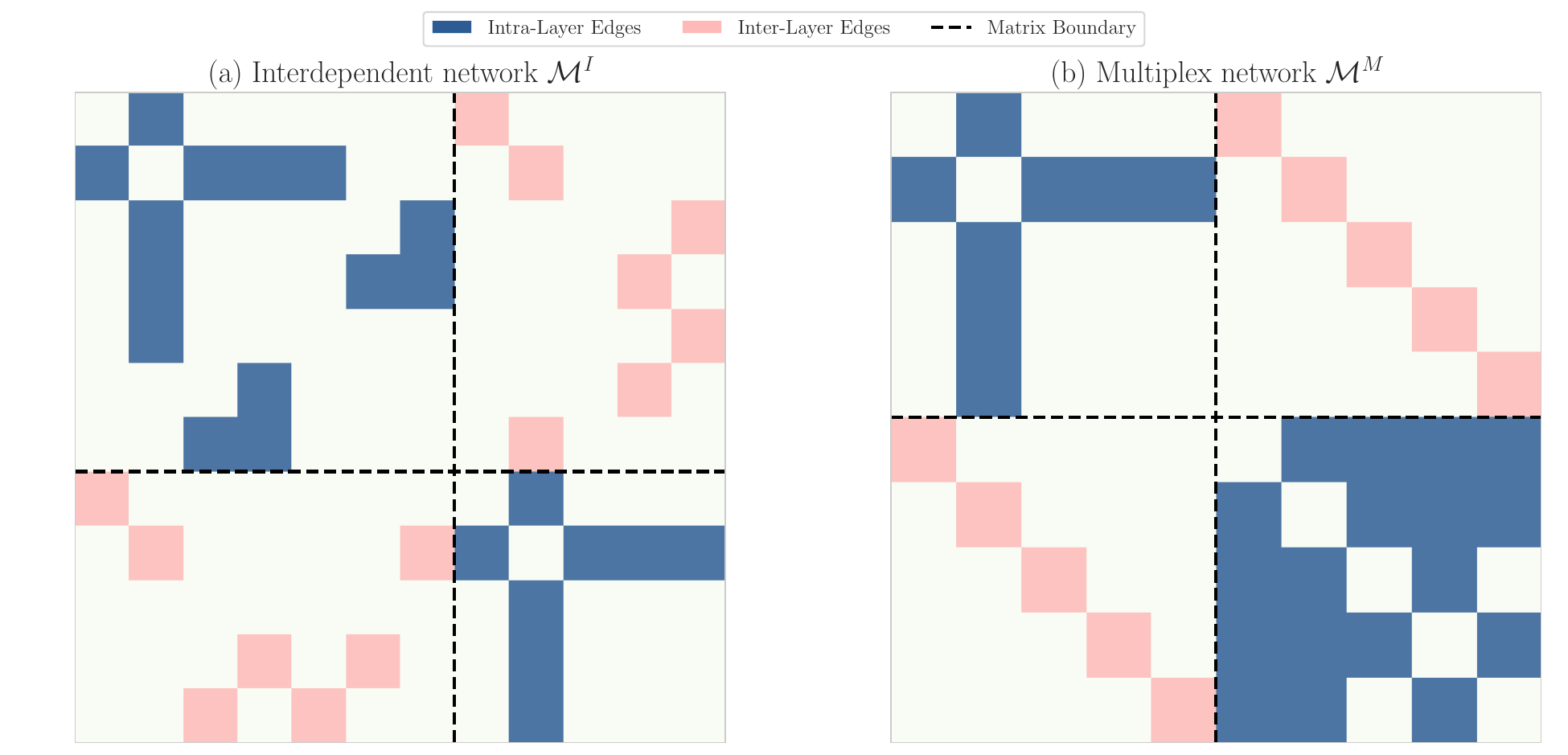}
	\end{center}
	\caption{Supra-adjacency matrices for the interdependent toy network $\mathcal{M}^I$ (left) and the multiplex toy network $\mathcal{M}^M$ (right). The networks are visualised in Figure \ref{fig:mlntypes}.}
	\label{fig:sa}
\end{figure}

The supra-adjacency matrices for two of the toy networks (interdependent $\mathcal{M}^I$ and multiplex $\mathcal{M}^M$) are visualised in Figure \ref{fig:sa}.

The coupling blocks (matrices) in multiplex networks are square and diagonal, while in interdependent networks this is not generally the case and the dimension of the coupling matrices is $N_1\times N_2$ if there are $N_1$ nodes in layer 1 and $N_2$ nodes in layer 2.

Another method of simplification is the aggregation of layers, where the adjacency matrices of all layers are summed to arrive at the aggregated adjacency matrix of size $N\times N$: $\mathbf{A}=\sum_\alpha A^\alpha$. Although in this case the information on different types of relationships between nodes is present in an aggregated manner, the complex structure is lost, and the analysis is the same as for monolayer networks. Common terms for such a representation include super-network, aggregated network, and projection network. A comprehensive review of MLN simplification techniques is presented in \cite{interdonato2020multilayer}.

Methods and metrics for multilayer networks are typically generalisations of methods on monolayer networks. Common concepts from network theory, such as node centrality, clustering coefficients, and shortest paths, are among the most researched metrics on MLNs. On the other hand, there are quantities specific to multilayer networks that are meaningless for monolayer networks, such as interconnectedness or similarity of layers. In the following subsections, we synthesise the most important metrics for MLNs.

\subsection{Metrics for multilayer networks}
\label{sec:metrics}

Similarly as for monolayer networks, we consider metrics at different scales of interaction: global, local, and intermediate-scale at the intersection of the former two levels. Many measures for MLNs are generalisations from monolayer networks; however, there is often more than one possible generalisation, as we are transferring the concepts into higher-dimensional structures. In addition, there are metrics that are specific to MLNs and are not sensible for monolayer networks.

Multiplex networks are a particularly prominent category of MLNs and are distinct from other categories in that the nodes in all layers represent the replicas of the same physical entities. Thus, there are some measures that are specific to multiplex networks. We will point to these distinctions where appropriate.

Several metrics are based on shortest paths, similar to monolayer networks. Finding the shortest paths on multilayer networks requires making some preliminary assumptions, among which a key one is whether travelling on inter-layer edges has a cost associated with it \cite{kivela2014multilayer}. In multiplex networks, the inter-layer couplings are categorical, meaning that they connect identical nodes, and assigning a cost to them is not sensible. On the other hand, in interdependent networks, a cost can often be naturally assigned to inter-layer edges. For example, if a MLN represents a multi-modal bus-metro PT system, the cost of an inter-layer edge might represent the distance or walking time between a bus stop and a metro stop represented by the respective nodes. In multiplex network the shortest paths problems are usually solved on an aggregated network, while in interdependent networks it is usually done on a flattened network, corresponding to its supra-adjacency matrix. Intra-layer edges may be weighted or unweighted, similarly as in the monolayer case.

\subsubsection{Local measures}

The treatment of local properties at the node level (and sometimes edge level) in monolayer networks is typically connected to determining relative node importance within the network, as reflected in various centrality measures. This is similar for multilayer networks where centrality definitions are adapted and generalised from monolayer network definitions. An important distinction between multiplex and interconnected networks is that in multiplex networks we get a vector of centralities for each node, corresponding to its importance in each of the layers. To get a scalar measure of centrality the scores from all layers need to be aggregated; usually, they are summed over. Some of the key references on node centrality in multilayer and multiplex networks are e.g., \cite{sole2016random}, \cite{wan2022identification}, \cite{battiston2014structural}. Below, we provide a list of the most commonly used node importance metrics in the multilayer context which are relevant for PT research.

\begin{itemize}
    \item \textbf{Degree}: in a monolayer network, the node degree measures the number of direct neighbours of the node; $k_i=\sum_{j}a_{ij}$, where $a_{ij}$ are the elements of its adjacency matrix. In multilayer networks, it is defined analogously, where the sum is over the respective row in the supra-adjacency matrix: 
    \begin{equation}
        k_i=\sum_{j}[\mathbf{A}]_{ij}.
    \end{equation}
    This definition includes both the intra-layer and inter-layer edges.
    \item \textbf{Multiplex degree}: in a multiplex network, a node generally has a different number of neighbours in each layer, and its degree is represented by a vector of intra-layer degrees:
    \begin{equation}
        \mathbf{k} = (k_1, ..., k_{\alpha}).
    \end{equation}
    \item \textbf{Overlapping degree} (multiplex): For an aggregated metric, overlapping degree $o_i$ is used, which is defined as the sum, or some other aggregation of the multiplex degree vector:
    \begin{equation}
        o_i = \sum_{\alpha}k_i^{\alpha}.
    \end{equation}
    \item \textbf{Shortest paths-based centralities}: 
    In monolayer networks, closeness centrality measures the inverse of the sum of distances of shortest paths between node $i$ and all other nodes. In MLNs, paths within a single layer as well as paths over several layers that include travelling on inter-layer edges must be considered. This is similar for betweenness centrality which measures the proportion of shortest paths that pass node $i$ relative to the number of all shortest paths \cite{sole2016random}.
    \item \textbf{Eigenvector-like centralities}: there are several generalisations of eigenvector centrality to multilayer networks. The formulations make use either of the supra-adjacency matrix or the adjacency tensor to determine relative node importance based on both intra-layer and inter-layer edge structure. In the case of the supra-adjacency matrix, the procedure is the same as for monolayer networks, and an eigenproblem is solved for $\mathbf{A}$. Mathematically, there are several versions of the eigenproblem for tensors, so there are several choices for the formulation of the eigenvector centrality in multiplex networks. In the case of multiplex networks, the obtained measures must be aggregated over all replicas of the same physical node. We refer to some key references for eigenvector-type centralities in multilayer networks: \cite{tudisco2018node, curado2021identifying, sola2013eigenvector, frost2024generalized, bergermann2022fast}. A distinct measure of eigenvector-like centrality is PageRank centrality which has been adapted for MLNs (e.g., \cite{halu2013multiplex, iacovacci2016functional, tortosa2021algorithm}).
    \item \textbf{Clustering coefficient}: measures the presence of triangles in the network, i.e., the tendency of nodes that have a same neighbour to also be connected. In monolayer networks it is defined as:
    \begin{equation}
        C_i = \frac{1}{k_i(k_i-1)}\sum_{j,k}a_{ij}a_{jk}a_{kj}.
    \end{equation}
    This is similar in MLNs, but there are again several possibilities: clustering (triangles) within a layer or inter-layer triangles.
    \end{itemize}

    In addition to the generalised standard measures, there are several that measure a node's or edge's role across layers and are thus specific to multiplex networks. Here, we list some of the most important metrics. Among the key references are e.g. \cite{battiston2014structural, de2015ranking}.
    
    \begin{itemize}
    \item \textbf{Edge overlap}: measures the presence of an edge across all layers:
    \begin{equation}
        o_{ij} = \sum_{\alpha} a_{ij}^{\alpha}.
    \end{equation}
    \item \textbf{Interdependence:} $\lambda_i$ is a metric based on shortest paths that measures the proportion of shortest paths that cross more than one layer \cite{morris2012transport, nicosia2013growing}. Measured at node-level, it is defined as:
        \begin{equation}
            \lambda_i = \sum_{j\ne i} \frac{\psi_{ij}}{\sigma_{ij}},
        \end{equation}
    where $\psi_{ij}$ is the number of shortest paths between nodes $i$ and $j$ that include edges on more than one layer, and $\sigma_{ij}$ is the number of all shortest paths between $i$ and $j$. The global metric is the average interdependence $\lambda = \sum_i \lambda_i/N$; $\lambda \in [0,1]$, where higher values mean that the more shortest paths cross more than one layer and the layers are more highly interdependent.
    \item \textbf{Participation coefficient}: quantifies the relative connectedness of a node across layers \cite{nicosia2015measuring}:
    \begin{equation}
        P_i = \frac{L}{L-1} \left[ 1 - \sum_{\alpha=1}^L \left( \frac{k_i^{\alpha}}{o_i} \right)^2 \right]
    \end{equation}
    where $k_i^{\alpha}$ is the degree of node $i$ in layer ${\alpha}$, and $o_i$ is the overlapping degree of node $i$.  $P \in [0,1]$ and $P_i=0$ if node $i$ is active in only one layer, and is closer to 1 the more similar its degree across all layers.
    \item \textbf{Node versatility}: most generally, information on node importance across layers can be aggregated into node versatility as \cite{de2015ranking}:
        \begin{equation}
        V_i = \sum_{\alpha} w_{\alpha} \, C_i^{\alpha},
        \end{equation}
    where $C_i^{\alpha}$ is a selected centrality measure of node $i$ in layer $\alpha$, and $w_{\alpha}$ is a weight of layer $\alpha$.
\end{itemize}

\subsubsection{Global measures}

The metrics at the global scale describe the behaviour of the multilayer network as a whole. Several widely-used metrics are adapted versions of monolayer network measures, among them average shortest paths (and related metrics such as efficiency and diameter), assortativity, and clustering.

Shortest path-based metrics
describe the well-connectedness of the network. The extension from monolayer to multilayer networks is straightforward; the main difference is that in multilayer networks the traversal of inter-layer edges is allowed (e.g. \cite{brodka2011shortest, ghariblou2017shortest}).

\begin{itemize}
    \item \textbf{Efficiency}: measures the ease of communication by considering the shortest paths between all pairs of nodes in the network. The definition is the same as for monolayer networks, the only difference is that here we must allow for travelling both within a layer and between layers:
    \begin{equation}
    E = \frac{1}{N (N - 1)} \sum_{i \neq j} \frac{1}{d^{\alpha\beta}_{ij}}
    \end{equation}    
    Efficiency in MLNs has been considered in e.g. \cite{noschese2024communication, noschese2024enhancing}.
    \item \textbf{Assortativity}: in monoplex networks, assortativity measures the tendency of the nodes to connect to other nodes that are similar in some respect, most often meaning they have a similar degree. Generalisation to multilayer networks is possible in several ways: assortativity within a single layer, or assortativity of layers in multiplex networks, also called \textbf{inter-layer degree correlation}\cite{nicosia2015measuring}, \cite{de2016degree}.

    \item \textbf{Modularity}: measures the network's structure in terms of densely connected groups of nodes \cite{mucha2010community, zhang2017modularity}.

\end{itemize}

\subsubsection{Intermediate-scale measures}

Mesoscale properties in monolayer networks act as an intermediate level between node-level (microscale) and global level (macroscale). In monolayer networks, the most common mesoscale object is a community, i.e., a group of nodes that act as a functional whole, and are at the same time different enough from other groups of nodes. In MLNs however, mesoscale can be understood in two distinct ways. The first is the generalisation of the concept of community and similar structures from monolayer to multilayer networks. Another class of mesoscale properties, specific to MLNs, are metrics that capture layer-level properties. In this case, a layer is understood as the group of nodes under consideration. There is a distinct class of information theory-based metrics that is specific to multilayer networks and is used to measure the structural (dis)similarity between layers.

\begin{itemize}
    \item \textbf{Community detection} in MLNs is a widely researched topic and there are many methods for multilayer or multiplex community detection (e.g. \cite{mucha2010community}, \cite{hamed2024comprehensive} \cite{boukabene2025community} \cite{kao2018layer}). 
    \item \textbf{Core-periphery detection}: examines the structure from the perspective of identifying the most central (i.e., core) part of the network \cite{bergermann2025core}.
    \item \textbf{Motif analysis}: motifs are recurring smaller structures within a network, e.g., triangles or some other special small subgraphs \cite{milo2002network}. This has been extended to multilayer networks \cite{battiston2017multilayer}, \cite{li2023multiplex}.
    \item \textbf{Layer similarity measures} such as \textbf{relative entropy} or \textbf{Jensen-Shannon divergence} quantify layer similarity and are often used to determine the optimal number of layers \cite{de2015structural}.  A comprehensive review of (dis)similarity measures is presented in \cite{brodka2018quantifying}.
\end{itemize}

\section{Literature review}
\label{sec:litrev}

We perform a structured literature review following the search process as described in Section \ref{sec:search} below.
The systematic review of the literature reveals distinctive categories of research topics with respect to transport problems addressed, which can be primarily grouped into three categories that are discussed in the respective subsections of this Section: $(i)$ network structure in \ref{sec:structure}, $(ii)$ PT resilience in \ref{sec:resilience} and $(iii)$ planning and optimisation problems in \ref{sec:planning}. We note that the categories are not classes at the same conceptual level, but are rather constructed bottom-up from the surveyed literature. In the following, we intentionally bypass any dedicated discussion on multilayer representations as this is the focus of Section \ref{sec:tax} where we provide a taxonomy of MLPTN representations and the many challenges related to it.

\subsection{Search process and survey results}
\label{sec:search}

First, we searched the databases WoS and Scopus. The query was: \texttt{("multilayer network*" OR "multi layer network*" OR "multiplex network*" OR "interdependent network*" OR "interconnected network*" OR "layered network*" OR "network of networks" OR "coupled network*" ) AND ("public transport" OR "public transit" OR "mass transit" OR "urban transit" OR "urban transport" OR "bus network*" OR "tram network*" OR "metro network*" OR "subway network*" OR "underground network*" OR "rail transit" OR "light rail" OR "commuter rail" OR "rapid transit" OR "public transport network*")}. The search was for title, abstract and keywords. We last performed the search on January 19 2026. We got 366 hits on Scopus and 94 on WoS.
After reading the abstracts and full texts, and snowballing in both directions, the set of the papers included in the review includes 63 articles. 

The criteria for inclusion (or exclusion) are:

\begin{itemize}
    \item The article focuses on public transport systems. Articles that do not consider PTNs are excluded. Works where public transport is added as a layer in a wider network (e.g., social or contact networks in epidemic spreading studies) are also excluded. We include in the review articles that have the development of new methods as their primary aim but where the application to PT problems is highly relevant also from the transport research perspective.
    \item In the case of multimodal systems, all modes must be PT. An exception to this point are papers where the main focus is on PT characteristics and inclusion of other modes does not compromise this focus.
    \item Networks must explicitly be modelled as multilayer networks; works where PTNs are modelled as ordinary networks are excluded. Works where the monolayer representation is a result of explicitly defined multilayer network aggregation or projection are included in the review.
    \item Only journal articles in English language are included. An exception is a conference publication \cite{shanmukhappa2018multi} that represents an early contribution and is often cited in subsequent related work. Conference or non-academic publications and non-English articles are not considered in the survey.
\end{itemize}

Note that, with regard to the first point, there is a substantial body of literature that considers multi-modal mobility, where all modes - PT, car, active modes, shared mobility, etc. - are included as layers in MLN representations of these networks (e.g. \cite{curado2021understanding}, and see \cite{alessandretti2023multimodal} for a recent review on multilayer multi-modal urban mobility). In this work, our focus is on different possibilities of modelling PTNs as mulilayer networks, therefore these works fall out of the scope of this survey.

Regarding the third point, we note that while we exclude articles where the PTNs do not exhibit an explicit MLN formulation, it is not possible in a systematic manner to include works where a multilayer structure is implicit and methods employed may well fall under the MLN methodology, yet are not described as such (e.g., \cite{yap2019shall, luo2020can}. 

The surveyed works are detailed in Tables \ref{tab:review1} - \ref{tab:review3}.

\subsection{Network structure}
\label{sec:structure}

Studies that consider network structure form a large part of the surveyed literature. Understanding structural, or topological, properties of PTNs and their relationship to system properties, such as resilience or accessibility, is a key research direction in the area of public transport networks. Among the 25 surveyed works in this group, we identify two distinct classes of studies. In the first, and larger group (17 articles), are the works that consider the classical metrics, such as node centrality, shortest paths, and global efficiency metrics are adapted to multilayer networks, and their multilayer generalisations are studied, primarily to understand the behaviour of multi-modal PT systems. The second category comprises 8 works that consider multilayer-specific metrics, examining the interaction between layers or layer similarity. The works in both categories are summarised in Tables \ref{tab:review1} and \ref{tab:review2}, respectively. Some articles consider both extended classical and multilayer-specific metrics, and we have classified them according to the primary focus. We point to these works in the text.

\subsubsection{Extended classical topological metrics}

Works in this group predominantly consider standard local and global topological metrics. Of 16 studies, 11 focus on identifying the most central nodes, and 7 consider global topological metrics such as efficiency or clustering; 4 of those consider both local and global metrics. In two articles, the focus is on identifying the shortest paths in MLPTNs.

\begin{table}[H]
\centering
\footnotesize
\caption{\footnotesize Overview of the works included in the systematic review - classical topological metrics extended to multilayer networks. PT modes: M - metro, B - bus, R - rail.
}
% \scalebox{0.75}{ 
\begin{tabular}{p{0.75cm}p{4.5cm}p{1.8cm}p{1.2cm}p{0.75cm}p{4.5cm}}
\toprule
\textbf{Study} & \textbf{Aim} & \textbf{Region} & \textbf{PT modes} & \textbf{Other modes} & \textbf{Main findings w.r.t. MLN perspective} \\
\midrule
\cite{kurant2006layered} & Comparing the topologies of infrastructural and passenger flow representations. & Warsaw, Switzerland, Central Europe & BMR & No & Topologies of both layers differ significantly, implying the need for considering both layers in estimating realistic loads. \\
\cite{alessandretti2016user} & Introducing a new multilayer representation from the users' perspective to model total travel time and assess PT system efficiency. & Greater Paris, Toulouse, Nantes, Strasbourg & BMR & No & The representation where each layer is a line in P-space (i.e. a clique) and inter-layer edge weights represent transfer waiting times offers a reliable method of estimating PTN efficiency from the users' perspective incorporating different components of travel time. \\
\cite{feng2017weighted} & Estimating spatiotemporal characteristics of both train and passenger flow distributions. & Beijing & M & No & Multilayer modelling of train and passenger flows is able to describe the differences in flows in both layers. \\
\cite{shanmukhappa2018multi} & Identifying central nodes based on centrality and land use. & London & BM & No & Different node importance scores in monolayer vs. multilayer representations; definition of supernodes based on spatial aggregation. \\
\cite{wu2019three} & Assessing transport efficiency and train usage for a metro system. & Tokyo, Hong Kong, Beijing & M & No & A model of passenger flow dynamics considering infrastructure and transfers. \\
\cite{zhang2020properties} & Assessing the performance and observing the evolution of the China High-Speed Rail network based on layers corresponding to service operations at different design speeds. & China & R & No & A method to assess overall network performance based on the service operation evolution as ever more connections are being serviced by high-speed routes. \\
\cite{wang2021network} & Analysing the properties of the Chinese high speed rail network from complementary perspectives of physical infrastructure and operation. & China & R & No & The infrastructural and operational layer have markedly different topologies and reflect  spatial and socioeconomical factors, respectively. \\
\cite{bergermann2021orientations} & Determining spatial orientation of PTNs and identifying the most central nodes in coupled PTNs. & 36 German cities and 18 European cities & Various & No & Formulation of matrix-based multiplex centralities to determine key nodes and lines in multi-modal PT systems. \\
\cite{tang2021identifying} & Fusing the data from metro and bus PTNs to identify the critical nodes in an interdependent multi-modal PTN. & Shenzhen & BM & No & MNDS metric better identifies critical nodes in a metro-bus multimodal PTN than conventional metrics. \\
\cite{gu2022using} & Assessing the scaling laws of population with multilayer centrality metrics. & Shanghai & BM & No & Population scales better with multilayer centrality metrics than those of a single-mode network. \\
\cite{pu2022topology} & Multilayer formulation of a multi-modal PT system and analysis of its topology. & Lanzhou & BR & No & Identification of key nodes in the interconnected system. \\
\cite{li2023understanding} & Comparing standard network-based metrics of PT networks for single layer and multiplex representations. & Shenzhen & BM & Shared bike & Cooperation between layers improves capacity and efficiency of multi-modal transport systems. \\
\cite{li2023clustering} & Clustering of metro nodes based on both structure and flows. & Nanjing & M & No & A method to group nodes based on infrastructure and flows. \\
\cite{zeng2024entropy} & Defining entropy-based node importance metrics for coupled networks that reflect both local and global information, and assess their value for resilience analysis. & Chengdu & BM & No & Different effects of different centrality measures on different resilience metrics and moderate to no improvement after entropy enhancement. \\
\cite{sun2024construction} & Identifying the most influential nodes in the coupled bus-metro network based on transfers. & Xi’an & BM & No & Different centrality distributions in single and coupled networks; bus network (larger) has the most influence on centrality in the coupled network. \\
\cite{zhou2025node} & Determining node importance by combining information from the coupled bus-metro system and land use. & Beijing & BM & No & Different nodes are identified as the most important in single-mode vs. coupled networks. \\
\cite{he2025identification}  & Enhancing the granularity in identifying the most important nodes based on k-shell decomposition in multilayer networks & Harbin & BM & No & An improved metric able to better identify critical nodes in the multilayer network as demonstrated by robustness assessment in targeted attacks. \\
\bottomrule
\end{tabular}
% }
\label{tab:review1}
\end{table}

Most studies consider \textbf{node centrality} (\cite{kurant2006layered, feng2017weighted, shanmukhappa2018multi, zhang2020properties, pu2022topology, gu2022using, zeng2024entropy, zhou2025node, bergermann2021orientations, he2025identification, zhou2025structural, sun2024construction}) or \textbf{global topological metrics} such as efficiency, average shortest paths, diameter and clustering coefficient \cite{wu2019three, pu2022topology, gu2022using, zhou2025structural, sun2024construction, li2023clustering, li2024investigating}. Several studies report that the most important nodes in the networks according to node centrality metrics differs significantly for monolayer and multilayer representations of the same network (e.g. \cite{li2024investigating}, \cite{tang2021identifying}, \cite{sun2024construction}, \cite{zhou2025node}). In addition, multilayer centrality measures correspond better to land use patterns (\cite{shanmukhappa2018multi}), identify critical nodes in robustness assessments (\cite{he2025identification, zeng2024entropy}) and scale better with population (\cite{gu2022using}) than their counterparts in monolayer networks. Similarly, centrality metrics in different layers often amount to different ranking of nodes in each layer (\cite{wang2021network}). Conversely, there is some evidence to suggest that nodes that exhibit the highest importance in single layers also tend to have the highest overlap across layers \cite{zhou2025structural}. This indicates the need for additional research into these questions and identifying possible causes for discrepancies, i.e. studying the impact of MLPTN modelling and the choice of metrics on the results obtained and then expand those to examine differences between cities. At the intermediate scale, \cite{li2024investigating} perform consensus community detection on a multilayer multi-modal PT system in Shenzhen to find that clusters based on passenger demand across all modes correspond well to the urban functional zones.

\textbf{Shortest paths} are the basis for calculating the efficiency and some centrality measures, such as closeness and betweenness centrality. In some studies, the analyisis of shortest paths presents a primary focus, formulating the notion of shortest paths in multilayer multi-modal systems \cite{gallotti2014anatomy} or modelling the shortest paths from the users' perspective by considering different parts of journeys (legs, transfers) in separate layers \cite{alessandretti2016user}.

This body of literature mirrors the early development of monolayer and single mode PTN studies in pioneering efforts to develop formalisms for identifying the critical nodes in multilayer representations of multi-modal networks.

\subsubsection{Multilayer-specific metrics}

A smaller body of literature considers the effects of the multilayer structure on the description of PT systems by analysing multilayer-specific metrics. These studies mainly focus on either \textbf{layer interaction} or \textbf{layer similarity}.

\begin{table}[H]
\centering
\footnotesize
\caption{\footnotesize Overview of the works included in the systematic review - multilayer-specific metrics. PT modes: M - metro, B - bus, R - rail, T - tram, F - ferry, A - air, W - water.
}
% \scalebox{0.75}{ 
\begin{tabular}{p{0.75cm}p{4cm}p{1.8cm}p{1.2cm}p{0.75cm}p{4cm}}
\toprule
\textbf{Study} & \textbf{Aim} & \textbf{Region} & \textbf{PT modes} & \textbf{Other modes} & \textbf{Main findings w.r.t. MLN perspective} \\
\midrule
\cite{gallotti2014anatomy} & Finding the shortest paths in multimodal PT systems where components of travel time are taken into account, and examining the role of each mode in contributing to such time-respecting shortest paths. & Great Britain & BMTRFA & No & Multimodality and the role of each mode dependent on geographical scale of travel. \\
\cite{aleta2017multilayer} & Examining the difference between two multilayer representations where either each line is a layer or each mode is a layer. & 9 European cities & BMT & No & Per line and per mode representations give complementary insights into PT systems. \\
\cite{zheng2018understanding} & Identifying major factors that influence coupling between bus and metro systems. & Shenzhen & BM & No & Multimodal trips are favoured for long trips; speed ratio not a strong predictor of coupling. \\
\cite{sui2019public} & Examining interaction between supply and demand in PTNs. & Chengdu and Qingdao & B & No & Developed a method for assessing quality of supply compared to demand based on layer structure comparison. \\
\cite{li2024investigating} & Assessing the structure and accessibility of the multi-modal transport system. & Shenzhen & BM & Shared bike, taxi & Introducing an original multiplex representation of multi-modal transport systems by dividing the city into high-granularity grid structure and take the cells as the set of physical nodes in the multiplex network. \\
\cite{zhou2025structural} & Examining the structure and reliability of multilayer PTNs. & Chengdu & BM & No & Most central nodes in single layers also tend to have largest overlap; high edge overlap increases robustness. \\
\cite{zhang2025should} & Finding the optimal number of layers to retain maximum information based on both network topology and transport system properties. & Chengdu and Yangtze River Delta Urban Agglomeration & BMTRWA & No & Layer aggregation results in significant information loss and retention of all layers is recommended. Topological metrics alone do not sufficiently measure information loss. \\
\cite{wang2025multimodal} & Fusing multi-modal data for network flow prediction. & Hong Kong & M & No & A model for forecasting passenger in- and out-flows. \\
\bottomrule
\end{tabular}
% }
\label{tab:review2}
\end{table}

The interaction between layers is studied in \cite{zheng2018understanding}, where couplings between layers in interdependent networks are examined.
In \cite{aleta2017multilayer} the authors compare different multilayer formulations of multimodal PTNs - each mode is a separate layer or each line is a separate layer - and find that the two representations offer complementary insights into PTN properties.

\cite{gallotti2014anatomy} and \cite{li2023understanding} examine the role of different modes in a multi-modal PT system where each layer in the MLN representation is a separate mode. They find that the different roles of distinct modes, such as metro carrying large passenger flows, and the bus system offering greater spatial coverage, are observed also in the MLN representation as differences between topological structures of layers. The potential merit of MLN modelling of PTNs for assessing the suitability of service supply to cater to passenger demand is studied in \cite{sui2019public}. There, one layer represents the network infrastructure and the other passenger flows. The representation is found to offer a tool for a comparative analysis of quality of PT systems in different cities.\cite{zhang2025should} investigated the relative information loss when progressively aggregating the fully multilayer representation of a multi-modal PTN to its monolayer network projection counterpart, using information-theoretic metrics. They conclude that any aggregation results in a considerable information loss, highlighting the need to keep the full multilayer structure for a comprehensive understanding of the system's behaviour. A different perspective is studied in \cite{wang2025multimodal}, where it is found that fusing data from different layers amounts to improved passenger flow predictions.

\subsection{Resilience analysis of MLPTNs}
\label{sec:resilience}

A substantial body of research has examined the resilience of PTNs, i.e., the extent to which transport systems can withstand, absorb, and adapt to disruptions.
While the general literature on PTN resilience is extensive, the subset of works that explicitly adopt multilayer network representations is considerably smaller. 
This focus is particularly important because disruptions evolve and propagate over time, affect the coupled behaviour of infrastructure, services, and demand, and may spread across different transport modes, all of which can be naturally represented within a multilayer network framework.
In this subsection, we focus on the works from the literature survey that study the resilience of MLPTNs. 

The surveyed works can be classified according to how they model disruptions, how they simulate the system response, and which performance indicators they evaluate.
Table~\ref{tab:resilience-categories} summarises the surveyed studies according to these classification dimensions, providing an overview of the type of disruption/recovery types, modelling assumptions, and performance indicators used in resilience analyses of MLPTNs.

\paragraph{Disruption-only analysis vs.\ disruption-and-recovery analysis.}
As shown in Table~\ref{tab:resilience-categories}, most studies focus exclusively on the effects of component failures---for example, how network connectivity or efficiency deteriorates under specific attack scenarios. These works typically evaluate robustness by simulating progressive removal of nodes, edges, or layers. In contrast, fewer studies---only 4 out of 32---also incorporate recovery processes. These approaches model the reactivation of components~\cite{yadav_resilience_2020, ju_multilayer_2022,zhong_restoration_2024}, rerouting services and demand~\cite{xu_modeling_2024}, or adaptive operational strategies~\cite{zhong_restoration_2024}, enabling the evaluation of resilience as a dynamic process that captures both performance loss and restoration.

\paragraph{Types of components affected: nodes, edges, or both.}
MLPTN robustness studies differ in how they represent disruptions. The majority---21 out of 32 studies---consider failures only at nodes (e.g., stations, stops, hubs), while only five studies simulate only edge failures (e.g., blocked track segments or road closures), and six works incorporate both node and edge disruptions, capturing more complex real-world scenarios. 

\paragraph{Types of attacks and disruption mechanisms.}
A wide range of failure scenarios has been explored:
\begin{itemize}

        \item \textbf{Isolated vs.\ cascading failures.} Most works model the (cumulative) failure of a set of components in the network, where each component fails independently (usually in a sequential manner). Other works address cascading effects, where a (usually single) initial failure triggers additional failures on other components due to overloading (e.g., ~\cite{zhao_cascading_2015,li_cascading_2024,zhu_congestion_2023}). MLPTNs are well-suited to represent both types of analysis and to assess how failures in one mode can directly impact dependent components in other layers.
    \item \textbf{Total vs.\ partial disruptions.} Many studies employ binary failure states in which components are either fully functional or fully disabled (e.g.,~\cite{yin_structural_2022,lu_measuring_2022}). Others consider partial failures, such as reduced capacity~\cite{ding_assessing_2025,ma_cascade_2023} or increased uncertainty in travel times~\cite{li_measuring_2024}, which provide a more nuanced representation of operational disturbances.
        \item \textbf{Random vs.\ targeted attacks.} Random disruptions are widely studied, as they approximate non-adversarial failures (e.g.,~\cite{baggag_resilience_2018,de_domenico_navigability_2014}).
        A special case is when the nodes/edges do not have a uniform probability of being selected but instead depend on external attributes such as exposure to extreme climate events (e.g.,~\cite{yadav_resilience_2020,zhang_extreme_2025}).
        Conversely, targeted attacks remove components according to topological indicators such as degree (e.g.,~\cite{ju_multilayer_2022,li_network-based_2018,hu_robustness_2025}), betweenness centrality (e.g.,~\cite{zhong_restoration_2024,liu_modelling_2020,xu_exploring_2024,liu_cascading_2022}), and/or to other related node/edge attributes such as passenger load (e.g.,~\cite{xu_modeling_2024,ren_analysis_2016}).   

\end{itemize}

\paragraph{Performance metrics and evaluation criteria.}
Throughout the simulated failure and recovery processes, resilience is assessed using a variety of metrics. Common topological indicators include the number of affected nodes/edges (e.g.,~\cite{li_cascading_2024,zhu_congestion_2023,hu_robustness_2025}), the size of the giant connected component (GCC) (e.g.,~\cite{palk_robustness_2026,zhang_can_2026,cao_robustness_2025,chen_evaluating_2024,liu_modelling_2020}), the average path length or diameter (e.g.,~\cite{li_network-based_2018, aparicio_assessing_2022}) or other functional indicators such as passenger overflow (e.g.,~\cite{zhu_congestion_2023,ding_assessing_2025,yang_temporal_2021,hu_robustness_2025}), system efficiency (e.g.,~\cite{zhao_cascading_2015,yin_structural_2022,gao_understanding_2025}), coverage ratio of a random walker~\cite{baggag_resilience_2018,de_domenico_navigability_2014}, accessibility measures~\cite{hong_vulnerability_2019}, or combined metrics based on the previous~\cite{ju_multilayer_2022}.

\medskip

The surveyed literature also reveals several methodological and conceptual challenges that remain open for future research, which are discussed in Section~\ref{sec:open-resilience}.

\begin{table}[!htbp]
\centering
\small
\caption{Summary of surveyed studies on MLPTN resilience. F/R denote if failure and/or recovery phases are studied; V/E indicate whether nodes and/or edges are affected; “type of disruption/recovery” specifies the attack/recovery strategy, and “metrics” lists evaluated indicators.}
\renewcommand{\arraystretch}{.95}
\begin{tabularx}{\textwidth}{
    p{0.4cm}   % ref
    p{.1cm}    % F
    p{.1cm}    % R
    p{.1cm}    % V
    p{.1cm}    % E
    p{6.5cm}   % type
    X          % metric
}
\toprule
\textbf{ref.} & \textbf{F} & \textbf{R} & \textbf{V} & \textbf{E} & \textbf{type of disruption/recovery} & \textbf{metrics}\\
\midrule

\cite{de_domenico_navigability_2014} & \ding{51} & \ding{55} & \ding{51} & \ding{55} & random & coverage of a random walker \\

\cite{yang_statistic_2014} & \ding{51} & \ding{55} &  \ding{51} & \ding{55} & random & number of failed nodes \\

\cite{zhao_cascading_2015}& \ding{51} & \ding{55} & \ding{51} & \ding{55} & random/targeted failure of one initial node + cascading disruptions  & efficiency \\

\cite{zhang_cascading_2016} & \ding{51} & \ding{55} & \ding{51} & \ding{55} & random initial failure & number of affected nodes (cascade effect)\\

\cite{ren_analysis_2016}& \ding{51} & \ding{55} & \ding{55} & \ding{51} & targeted (top 10 with highest load) & size of GCC, mean path length, diameter, and efficiency\\

\cite{li_network-based_2018} & \ding{51} & \ding{55} &  \ding{51} & \ding{55} & random and targeted(node degree) & size of GCC, average path length, diameter and efficiency \\

\cite{baggag_resilience_2018} & \ding{51} & \ding{55} & \ding{55} & \ding{51} &random & coverage of a random walker \\

\cite{hong_vulnerability_2019} & \ding{51} & \ding{55} &  \ding{51} & \ding{51} & random and targeted (node degree, betweenness and a real rainstorm) & accessibility to different opportunities \\

\cite{liu_modelling_2020} & \ding{51} & \ding{55} &  \ding{51} & \ding{55}& single node targeted (betweenness and passenger load) attack & size of GCC and affected nodes \\

\cite{yadav_resilience_2020} & \ding{51} & \ding{51} &  \ding{51} & \ding{55} & failures: random, flooding probability, targeted (degree and betweenness), and compound (flood+targeted). Recovery: random, greedy, and based on centrality (betweenness, degree, eigenvector) & size of GCC and network functionality based on efficiency \\

\cite{jo_cascading_2021} & \ding{51} & \ding{55} & \ding{51} & \ding{55} & random initial failure & number of affected nodes (cascade effect)\\

\cite{yang_temporal_2021} & \ding{51} & \ding{55} &  \ding{51} & \ding{55} & 1–-2 manually selected nodes & delay, passenger volumes \\

\cite{aparicio_assessing_2022}& \ding{51} & \ding{55} & \ding{51} & \ding{51} & random removal, targeted (node degree and betweenness), and entire line removed & average path length, average degree, isolated components, size of GCC\\

\cite{liu_cascading_2022}& \ding{51} & \ding{55} & \ding{51} & \ding{51} & random and targeted (degree, betweenness, load) & size of GCC, mean shortest path length \\

\cite{lu_measuring_2022}& \ding{51} & \ding{55} & \ding{51} & \ding{55} & targeted single node (degree, betweenness, and passenger volume) and full mode disruption & interdependency between the modes \\

\cite{ju_multilayer_2022} & \ding{51} & \ding{51} &  \ding{51} & \ding{55}& targeted (node degree) & combined metric (heterogeneity, correlated coefficient, average path length, and efficiency)\\

\cite{yin_structural_2022} & \ding{51} & \ding{55} &  \ding{51} & \ding{51} & individual failure of each node, edge, and line & efficiency and ratio of overloaded edges \\

\cite{ma_cascade_2023}& \ding{51} & \ding{55} & \ding{51} & \ding{55} & targeted (degree) with station resistance & number of failed nodes, average efficiency, Shannon entropy of node degree\\

\cite{zhu_congestion_2023}& \ding{51} & \ding{55} & \ding{51} & \ding{55} & targeted attack on one node (degree, passenger flow, largest initial state) + cascading effect & proportion of congested stations \\

\cite{li_cascading_2024}& \ding{51} & \ding{55} & \ding{51} & \ding{55} & targeted attack on one node (degree, passenger flow, number of coupled nodes and coupling strength) + cascading effect & number of failed nodes and mean degree\\

\cite{chen_evaluating_2024}& \ding{51} & \ding{55} & \ding{51} & \ding{51} & random attacks, targeted attacks (node degree, clusters based on POIs) & size of GCC\\

\cite{xu_exploring_2024}& \ding{51} & \ding{55} & \ding{51} & \ding{55} &targeted attack (degree and betweenness) & failed stations, network effectiveness\\

\cite{li_measuring_2024} & \ding{51} & \ding{55} &  \ding{55} & \ding{51} & uncertainty added to the travel time & route diversity \\

\cite{xu_modeling_2024} & \ding{51} & \ding{51} & \ding{51} & \ding{55}  & targeted (passenger flow) & efficiency (distance and travel cost)\\

\cite{zhong_restoration_2024} & \ding{51} & \ding{51} & \ding{51} & \ding{55} & Disruption: targeted (betweenness). Recovery: random, control of flow, and greedy (load) & failed stations and size of the GCC \\

\cite{zhang_extreme_2025}& \ding{51} & \ding{55} & \ding{55} & \ding{51} & random rainfall events causing station or line closure & network efficiency  \\

\cite{ding_assessing_2025}& \ding{51} & \ding{55} & \ding{51} & \ding{51} & complete and partial disruptions (lowered capacity, increased demand) & overflow (number of affected passengers)\\

\cite{cao_robustness_2025} & \ding{51} & \ding{55} &  \ding{51} & \ding{55} & random and targeted (degree, betweenness, coreness, efficiency, covered population and neighbors reachable without transfers) & size of the GCC and efficiency \\

\cite{hu_robustness_2025} & \ding{51} & \ding{55} &  \ding{51} & \ding{55}  & targeted (node degree) & overloaded links \\

\cite{gao_understanding_2025} & \ding{51} & \ding{55} &  \ding{55} & \ding{51} & random, all links within a grid cell, and targeted (connectivity loss) & unreachable nodes and efficiency \\

\cite{zhang_can_2026} & \ding{51} & \ding{55} &\ding{51} & \ding{55}  & random and targeted (degree, betweenness and passenger flow) & size of the GCC \\

\cite{palk_robustness_2026} & \ding{51} & \ding{55} &\ding{51} & \ding{55}  & random and targeted (degree, closeness, betweenness and PageRank) & size of the GCC \\

\bottomrule
\end{tabularx}
% \caption{Summary of surveyed studies on MLPTN resilience. F/R denote if failure and/or recovery phases are studied; V/E indicate whether nodes and/or edges are affected; “type of disruption/recovery” specifies the attack/recovery strategy, and “metrics” lists evaluated indicators.}
\label{tab:resilience-categories}
\end{table}

\subsection{Service planning and design}
\label{sec:planning}

Compared to network structure and resilience studies, the number of PTN planning and design works is considerably smaller, and we found only a handful of studies that explicitly consider multilayer network structure in their formulation of a related optimisation problem. Two of the seven identified studies consider the line planning problem, two consider synchronisation in PT systems, and the remaining three works consider optimal transport in multi-modal PT systems, optimal transfer costs in multi-modal systems, and finding shortest paths in a dynamic network, respectively. Articles addressing optimisation problems are summarised in Table \ref{tab:review3}.

\begin{table}[!htbp]
\centering
\footnotesize
\caption{\footnotesize Overview of the works included in the systematic review - optimisation. PT modes: M - metro, B - bus, T - tram.
}
% \scalebox{0.75}{ 
\begin{tabular}{p{0.75cm}p{4.5cm}p{1.5cm}p{1.2cm}p{0.75cm}p{4.5cm}}
\toprule
\textbf{Study} & \textbf{Aim} & \textbf{Region} & \textbf{PT modes} & \textbf{Other modes} & \textbf{Main findings w.r.t. MLN perspective} \\
\midrule
\cite{du2016physics} & Finding optimal transfer costs between two transport modes with different travel speeds. & Simulated & Any & No & A method for analysing coupling between layers and a metaheuristic algorithm to optimise cooperation between layers based on setting optimal transfer costs. \\
\cite{pu2017analysis} & Determining the conditions for synchronisation of PT lines based on a super-network representation of a single mode. & Simulated & Any & No & A novel MLPTN representation with P-space and C-space representations of the same PT system as heterogeneous node layers. \\
\cite{du2019synchronisation} & Studying synchronisation in multi-modal PT systems based on route-level interactions. & Simulated & BM & No & A model for synchronisation of multi-modal PT system line operation. \\
\cite{ibrahim2021optimal} & Developing an optimal transport model for multilayer networks considering different impacts of distinct modes on total cost. & Bordeaux & BT & No & Optimal transport is influenced by different factors in different layers in multi-modal PT systems. \\
\cite{henry2021reinforce} & Developing a method to prioritise construction of newly-built PT lines from an existing itinerary, applied also to emergency events. & Lyon & MT & Road & Fast identification of optimal additional lines in the network by calculating multilayer topological metrics. \\
\cite{canca2023multilayer} & Developing a multilayer network formulation and solving the line planning problem for a coupled bus-pedestrian network. & Seville & B & Walk & Multilayer network able to represent all parts of the journey and all components of travel time when considering line planning problems. \\
\cite{yang2024mathematical} & Developing an algorithm for finding shortest paths in static and dynamic multilayer networks. & Beijing & B & No & The community structure of the layers allows for a flexible shortest path finding in time-varying traffic conditions.\\
\bottomrule
\end{tabular}
% }
\label{tab:review3}
\end{table}

The line planning problem is addressed in two alternative ways. \cite{henry2021reinforce} develops a method for fast identification of the lines that should be prioritised when adding new lines from a predefined itinerary to the existing PTN. Their methodology is based on calculating several topological metrics, both local and global, in multilayer representations of multi-modal networks. In contrast, \cite{canca2023multilayer} formulate and solve the line planning problem on a multilayer representation of PT journeys, where each layer represents a separate component of the journey (in-vehicle, transfers, together with walking to and from the station).

The synchronisation of a multi-modal PTN is examined in \cite{du2019synchronisation} where each layer represents the C-space representation of a mode. A similar model, and with a higher level of detail, is presented in \cite{pu2017analysis}, with an original representation where one layer represents service lines (i.e., C-space) and the other layer corresponds to the P-space representation. The derived representation is a so-called supernetwork, a network of networks, where each line node contains the clique of the stop nodes of the line. The authors develop a method for PTN synchronisation that is able to simultaneously capture both stop- and line-level effects.

A different problem is presented in \cite{ibrahim2021optimal}, where the authors investigated the different roles of different transport modes by encoding them in a multilayer representation. Their approach enabled the differentiation of the effects of different modes on the total travel cost, and consequently the system efficiency. The results show that different modes have a different impact on optimal flows in the overall multi-modal network.

In an early work on a simulated network, the effects of coupling between two layers with different design speeds were studied \cite{du2016physics}. Optimising transfer costs was found to strengthen cooperation between layers. In addition to a simple degree-based cost setting, a particle swarm optimisation algorithm was employed to further optimise flows.

\cite{yang2024mathematical} is a rare example of a dynamic multilayer network. It proposed a method for finding shortest paths on multilayer networks with time-dependent inter-layer edge weights. Besides finding shortest paths, the authors propose a method for creating a multilayer network from a monolayer one by an algorithm similar to community detection; each community is then represented as a separate layer. The method is applied to the bus network in Beijing where the urban area is separated into 11 districts, forming 11 layers. Both intra- and inter-layer edges represent infrastructural connections, within the district and between districts, respectively.

\section{Taxonomy of MLPTNs}
\label{sec:tax}

\subsection{Proposed taxonomy}

Based on the survey, we identify several key aspects of multilayer public transport networks that in our view will benefit from conceptual clarification and categorisation.
A major critical observation is that the terminology used in the literature lacks a clear background and that the reported type of the representation used often does not agree with the established literature.
For example, several studies report that they model a PT system as a multiplex network, whereas those are in fact treated as interdependent networks (e.g., \cite{tang2021identifying}, \cite{zheng2018understanding}, \cite{li2023understanding}). In contrast to other categories proposed in Section \ref{sec:mlntypes} (interdependent, heterogeneous), multiplex networks are well-defined, and their correct classification is not merely a matter of naming, but has deeper implications, as several methods are developed specifically for multiplex networks and only make sense in this setting, see for example the designated metrics in \ref{sec:metrics}. The situation is less clear in the case of interdependent and heterogeneous networks, for which the terminology is not used universally and consistently in theory and across application domains. The classification we proposed in Section \ref{sec:mlntypes} is meant to address the issue of the multilayer representation used.

In this section, we address the issue of PTN-specific multilayer representations. From the surveyed literature, several recurring ways of modelling PTNs as multilayer networks are apparent. Of those, the most common are representing each mode of a multi-modal system as a separate layer, or representing each line in a single mode system as a separate layer. A fundamentally different representation takes one of the layers to encode infrastructure and another to represent OD flows.
However, there is no agreed-upon terminology for these representations. This often makes it unclear what the underlying reasons for a certain choice of representation are as well as hinders the comparison of findings across different studies.

\begin{figure}[H]
	\begin{center}
		\texttt{}\includegraphics[scale=0.425]{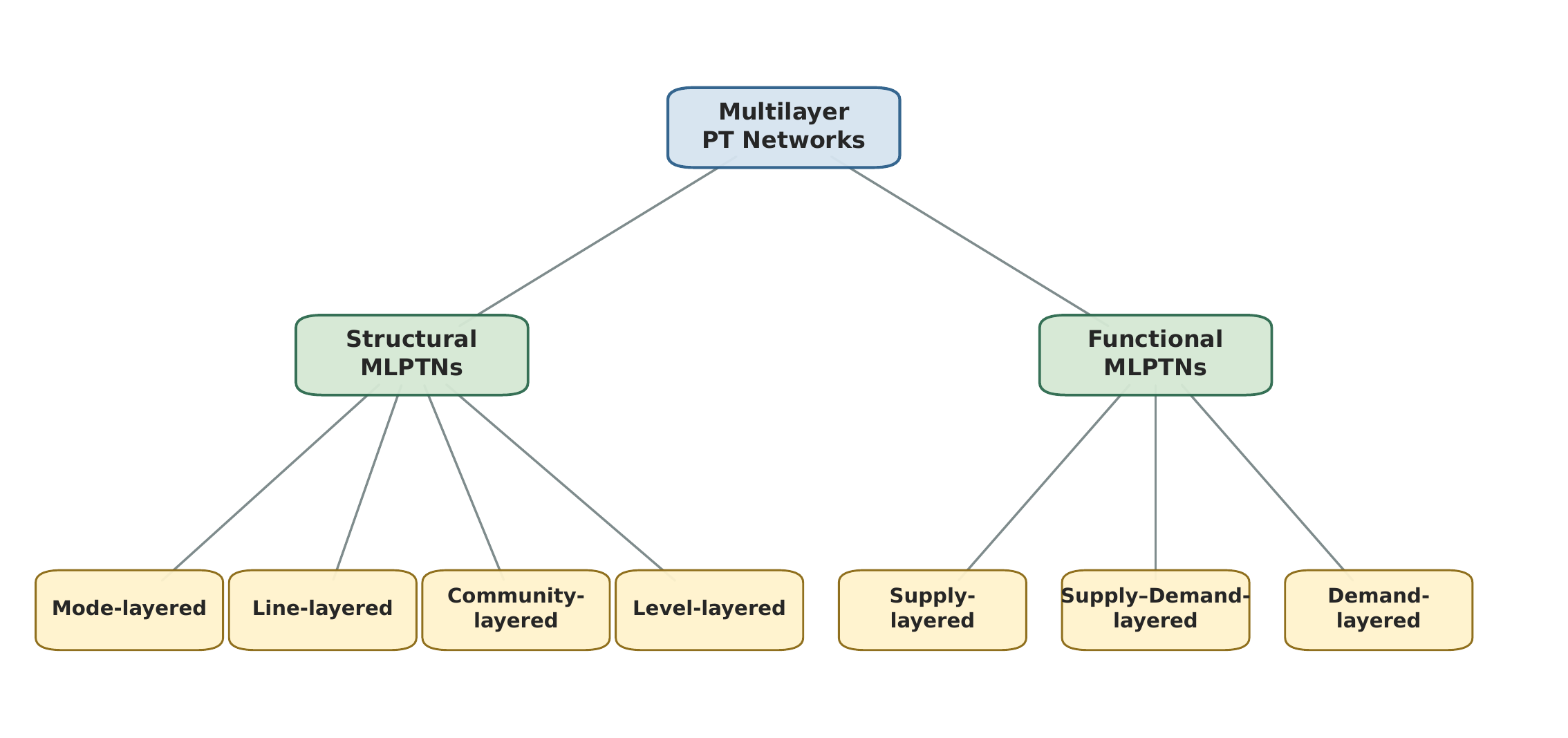}
	\end{center}
	\caption{A two-level taxonomy of MLPTNs.}
	\label{fig:mlntax}
\end{figure}

To clarify these inconsistencies and to gain a deeper understanding of the role of---and subsequently the potential for---using multilayer network representations in PT systems analysis, we propose a taxonomy of multilayer public transport networks. The taxonomy tree is shown in Figure~\ref{fig:mlntax}.

First, we note that representations can be divided into those that are layered based on PT (infra)structure and those that include layers which corresponds to distinctive functional notions.
% include only supply information and those that include both supply and demand aspects. 
This is our first level of distinction, and we classify MLPTNs as either \textbf{Structural MLPTNs} or \textbf{Functional MLPTNs}. Those are then further divided based on the guiding principle for separating the network into layers. Each of the two types is divided into several subtypes as detailed in the following.

Structural MLPTNs can be divided into: 
\begin{itemize}
    \item \textbf{Mode-layered}, where each layer represents a separate mode of a multi-modal PT system. These are typically (and most naturally) modelled as interdependent MLNs, as nodes correspond to similar notions (PT stations or stops). However, the actual stations of different modes represent different physical objects and may be characterized by different variables.
    \item \textbf{Line-layered}, where each layer represents a single PT line. Line-layered MLPTNs can be modelled as multiplex networks where the inter-layer edges connect nodes that represent the same stops. However, this can be extended to interdependent representations, where inter-layer edges include weights that represent waiting times, or can connect nodes representing stops within walking distance.
    \item \textbf{Community-layered}, where each layer represents a community of nodes, based on some similarity condition, e.g. geographical clustering. Intra-layer edges represent the infrastructural or operational connections between the nodes within the cluster (i.e., are subgraphs of L- or P-space representations), and inter-layer edges are again the infrastructural or operational edges, but connect respective nodes from different clusters. Community-layered MLPTNs can be modelled as interdependent or multiplex networks.
    \item \textbf{Level-layered}, where each layer represents a different level of aggregation of PT elements. An example is a representation where one layer represents the P-space and the other the C-space network representation of the same PTN. Level-layered MLPTNs are modelled as heterogeneous networks.
\end{itemize}

Functional networks contain layers that encode different functions of the PT system. Thus, they are most naturally modelled as multiplex networks, but we do not exclude other possibilities. We divide them into three subcategories:

\begin{itemize}
    \item \textbf{Supply-layered}: layers represent different aspects of supply, e.g., one layer can represent infrastructural connections and the other operational connections.
    \item \textbf{Supply-Demand-layered}: layers represent aspects of supply and of demand. For example, one layer is an infrastructural representation and the other captures OD flows.
    \item \textbf{Demand-layered}: captures different aspects of demand. An example is a multiplex network where one layer represents OD flows for whole journeys, another OD pairs for legs of journeys, and a separate layer represents transfers.
\end{itemize}

We emphasise that the classification is based on the decision of structuring the system into different layers, and has implications for the modelling of which as an MLPTN. The analysis may then proceed with other properties included in the network. For example, it may happen that edge weights in Structural MLPTNs represent passenger flows or vehicle occupancy. Although the weights reflect demand-related quantities, we classify this as Structural networks due to the nature of layer separation based on the (infra)structure of the PT system.

Distinct MLPTN representations in our taxonomy are often most naturally modelled with specific MLN types, as we indicate for each category in the above taxonomy. However, there is no one-to-one correspondence between the two classifications, and MLN and MLPTN categories offer two orthogonal axes of classification. Indeed, original combinations present an interesting future research direction, as we discuss in Section \ref{sec:agenda}.

\subsection{Classification of surveyed literature}

We apply the proposed taxonomy to the studies included in the Survey reported in the previous section. Table \ref{tab:mlptn.tax} notes for each study its classification in terms of the MLNs (Section 2.2) used as well as the proposed taxonomy of MLPTNs (section 4.1).

\begin{table}[H]
\centering
\footnotesize
\caption{Classification of related works according to the proposed taxonomy.}
\begin{tabular}{llll}
\toprule
Study & MLN type & MLPTN type & MLPTN subtype \\
\midrule
\cite{kurant2006layered} & Multiplex & Functional & Supply-Demand-layered \\
\cite{alessandretti2016user} & Interdependent & Structural & Line-layered \\
\cite{feng2017weighted} & Multiplex & Functional & Supply-Demand-layered \\
\cite{shanmukhappa2018multi} & Interdependent & Structural & Mode-layered \\
\cite{wu2019three} & Multiplex & Functional & Supply-Demand-layered \\
\cite{zhang2020properties} & Interdependent & Functional & Supply-layered \\
\cite{tang2021identifying} & Interdependent & Structural & Mode-layered \\
\cite{wang2021network} & Multiplex & Functional & Supply-layered \\
\cite{bergermann2021orientations} & Multiplex & Structural & Line-layered \\
\cite{gu2022using} & Interdependent & Structural & Mode-layered \\
\cite{pu2022topology} & Interdependent & Structural & Mode-layered \\
\cite{li2023clustering} & Multiplex & Functional & Supply-Demand-layered \\
\cite{li2023understanding} & Interdependent & Structural & Mode-layered \\
\cite{sun2024construction} & Interdependent & Structural & Mode-layered \\
\cite{zeng2024entropy} & Interdependent & Structural & Mode-layered \\
\cite{he2025identification}  & Interdependent & Structural & Mode-layered \\
\cite{zhou2025node} & Interdependent & Structural & Mode-layered \\
\cite{gallotti2014anatomy} & Interdependent & Structural & Mode-layered \\
\cite{aleta2017multilayer} & Multiplex & Structural & Line-layered \\
\cite{zheng2018understanding} & Interdependent & Structural & Mode-layered \\
\cite{sui2019public} & Multiplex & Functional & Supply-Demand-layered \\
\cite{li2024investigating} & Multiplex & Structural & Mode-layered \\
\cite{wang2025multimodal} & Multiplex & Functional & Demand-layered \\
\cite{zhang2025should} & Interdependent & Structural & Mode-layered \\
\cite{zhou2025structural} & Interdependent & Structural & Mode-layered \\
\cite{de_domenico_navigability_2014} & Interdependent & Structural & Mode-layered \\
\cite{yang_statistic_2014} & Interdependent & Structural & Mode-layered \\
\cite{zhao_cascading_2015} & Interdependent & Structural & Mode-layered \\
\cite{ren_analysis_2016} & Multiplex & Functional & Supply-Demand-layered \\
\cite{zhang_cascading_2016} & Multiplex & Functional & Supply-Demand-layered \\
\cite{li_network-based_2018} & Interdependent & Structural & Mode-layered \\
\cite{baggag_resilience_2018} & Interdependent & Structural & Mode-layered \\
\cite{hong_vulnerability_2019} & Interdependent & Structural & Mode-layered \\
\cite{liu_modelling_2020} & Interdependent & Structural & Mode-layered \\
\cite{yadav_resilience_2020} & Interdependent & Structural & Mode-layered \\
\cite{jo_cascading_2021} & Interdependent & Structural & Mode-layered \\
\cite{yang_temporal_2021} & Heterogeneous & Functional & Supply-Demand-layered \\
\cite{aparicio_assessing_2022} & Interdependent & Structural & Mode-layered \\
\cite{liu_cascading_2022} & Interdependent & Structural & Mode-layered \\
\cite{lu_measuring_2022} & Interdependent & Structural & Mode-layered \\
\cite{ju_multilayer_2022} & Interdependent & Structural & Mode-layered \\
\cite{yin_structural_2022} & Interdependent & Structural & Line-layered \\
\cite{ma_cascade_2023} & Interdependent & Structural & Mode-layered \\
\cite{zhu_congestion_2023} & Interdependent & Structural & Mode-layered \\
\cite{li_cascading_2024} & Interdependent & Structural & Mode-layered \\
\cite{chen_evaluating_2024} & Interdependent & Structural & Mode-layered \\
\cite{xu_exploring_2024} & Interdependent & Structural & Mode-layered \\
\cite{li_measuring_2024} & Heterogeneous & Functional & Supply-layered \\
\cite{xu_modeling_2024} & Interdependent & Structural & Mode-layered \\
\cite{zhong_restoration_2024} & Interdependent & Structural & Mode-layered \\
\cite{ding_assessing_2025} & Multiplex & Functional & Supply-Demand-layered \\
\cite{zhang_extreme_2025} & Interdependent & Structural & Line-layered \\
\cite{cao_robustness_2025} & Interdependent & Structural & Mode-layered \\
\cite{hu_robustness_2025} & Interdependent & Structural & Mode-layered \\
\cite{gao_understanding_2025} & Interdependent & Structural & Mode-layered \\
\cite{zhang_can_2026} & Interdependent & Structural & Mode-layered \\
\cite{palk_robustness_2026} & Interdependent & Structural & Mode-layered \\
\cite{du2016physics} & Interdependent & Structural & Mode-layered \\
\cite{pu2017analysis} & Heterogeneous & Structural & Level-layered \\
\cite{du2019synchronisation} & Interdependent & Structural & Mode-layered \\
\cite{ibrahim2021optimal} & Interdependent & Structural & Mode-layered \\
\cite{henry2021reinforce} & Interdependent & Structural & Mode-layered \\
\cite{canca2023multilayer} & Interdependent & Structural & Line-layered \\
\cite{yang2024mathematical} & Interdependent & Structural & Community-layered \\
\bottomrule
\end{tabular}
\label{tab:mlptn.tax}
\end{table}

It becomes evident from Table \ref{tab:mlptn.tax} that Structural MLPTNs dominate in the existing literature. By far the most common representation is the mode-layered MLPTN modelled as an interdependent network, followed by Line-layered MLPTNs. Among Functional MLPTNs, Supply-Demand-layered are the most common, where the comparison between network structure and passenger flows is studies \cite{sui2019public, li2023clustering}. In an early conceptual work \cite{kurant2006layered}, the layering of a PTN into an infrastructural (physical) and passenger flows (logical) layers was put forward, motivating subsequent works. Among the less-represented types, an interesting example of Supply-layered functional MLPTNs is presented in \cite{wang2021network}. Here, one layer represents rail infrastructure, and the other layer represents OD pairs of service lines, i.e., connects stations that are the endpoints of a line, where the latter differs from both L- and P-space representations. This representation allows for a novel comparison of topologies. Another is a Demand-layered representation in \cite{wang2025multimodal} where one layer represents in-flows and the other out-flows of passengers. \cite{pu2017analysis} model the PT system as a heterogeneous network, where one layer includes the L-space representation - a stop-level description -, and the other the C-space representation - a line-level description. This allows for a multi-level formulation of the synchronisation problem. An interesting approach is that in \cite{yang2024mathematical} where a single-mode large network (the bus system in Beijing) is restructured so that each of the 11 city regions represents a separate layer in a multilayer network. This allows for modelling varying travel times within each region separately.
In exceptional cases, proposed representations may fall in between the main categories identified in our taxonomy. 
One example is~\cite{li_measuring_2024}, where a line-layered network is initially modelled, with each layer later split into slices to account for supply across time, thus combining elements of structural and functional representations.

We discuss the under-representation of several MLPTN subtypes, and the potential for their use in research and planning, as one of the more promising topics for future work in the following section.

\section{Research agenda}
\label{sec:agenda}

Based on the results of our survey and the proposed taxonomy, we outline our vision for the research into multilayer public transport networks. We do so based on the following observations in relation to the current state of the art: $(i)$ there is a lack of unified terminology in MLPTN research and inconsistent adherence to the established terminology in theory and other application domains; $(ii)$ related to the previous point, the motivation for using multilayer representations is often unclear; $(iii)$ multilayer representations are, with a few exceptions, limited to the most obvious ways (such as for multimodal systems), and (partially as a consequence) $(iv)$ there is a disproportionate focus on certain topics, especially basic network statistics and robustness-related studies.

In the following, we address each of the above-mentioned shortcomings and propose a series of research questions which are meant to overcome those. In addition, we note that many observations may be valid also for other areas of transport.

\subsection{Unifying the terminology and consolidating the field}
The integration of public transport research and multilayer networks faces many of the same difficulties as the application of monolayer networks to PT research encountered in the past. This is further amplified by the increased complexity of MLNs compared to monolayer networks and opens new conceptual venues of investigation. In the following, we propose several research questions that may guide the integration.

\begin{itemize}
    \item How can existing metrics for multilayer networks be (re)interpreted within the public transport domain and how can they be modified to best reflect PT properties of interest to transport researchers and planners?
    \item How can the identification of relevant PT properties guide/inspire the development of new MLN metrics tailored for the analysis of MLPTN?
    \item What methods of data collection and processing can streamline the construction of MLPTN representations?
    \item How can multilayer networks with heterogeneous nodes be employed to explore the relations between conceptually different entities within a unified representation (e.g., relations between stations and routes, or stations, lines and passengers)?
    \item What functional relations between nodes should be meaningfully introduced as separate layers? 
    \item When is the multilayer representation advantageous over aggregated representations and what methods can we use to quantitatively identify the optimal representation?
    \item What types of node metric aggregation over layers can be devised to reliably reflect the desired PT system properties? 
    \item What is the potential of interconnected networks in studying the interdependency of transport systems at different scales (e.g., PT networks of several cities connected by interregional transport systems)?
    \item Can multilayer networks be used for the reconstruction/inference of missing information (e.g., OD flows from structure)?
\end{itemize}

In addressing these research questions it is crucial that future studies clearly specify the \textbf{Multilayer network type} (multiplex, interconnected network, ...) which are now often used as if they are interchangeable in the literature. Moreover, some terms include others (e.g., "multilayer network" encompasses "multiplex network", but not vice versa)---the most stringent definition should be used. This also makes it clearer which properties of the network are most important and which sets of methods and metrics are readily applicable. It should also be made clear what the basis is for separating nodes into layers (\textbf{Type}: Supply MLPTN, Supply-Demand MLPTN, ...) and what different layers represent (Mode-layered, Line-layered, ...), potentially inspired by the taxonomy presented in the previous section. In addition, the objective of the study should clearly state the transport problem it is addressing and how the specific choice is suitable for the analysis carried out.
Finally, related studies should clearly describe the \textbf{Intra-layer edge weights} adopted, the \textbf{Inter-layer couplings}---this is the least studied aspect and conceptual work is needed---, the \textbf{Mathematical formalism} employed (tensors, supra-adjacency matrix, aggregation ...), and the \textbf{Metrics} studied (centrality, layer similarity, ...).

\subsection{Representing and modelling of MLPTNs}

As is evident from our taxonomisation based on the literature, there are large variations in the amount of research effort which has been devoted to different MLN and MLPTN representations, and the related transport problems considered, with a large focus on a few most obvious representations. Some of the previously used representations are under-represented, and some are completely absent. While it may not make sense for each combination of MLN and MLPTN types to give a useful representation, we see vast uncovered potential in devising conceptually different MLPTN representations to guide transport research in different directions.

Besides focusing on topological representations, future research will best benefit transport planners by considering the spatial and temporal dimensions of PTNs. Public transport networks are essentially spatial networks \cite{barthelemy2011spatial}, however, their geographical properties are rarely taken into account in a comprehensive manner beyond including distances or travel times as edge weights. In addition, real-world PT systems are highly dynamic, with their properties varying over time: e.g., OD flows differ across hours in a day, or days in a week, as do service frequencies; in the longer run, new nodes or edges are added or existing ones removed. Temporal networks are a general term for time-varying networks. One of the common formulations sees them as a subcategory of multiplex networks, where consecutive layers represent static snapshots of a network at consecutive discrete time steps \cite{holme2012temporal}.

Below we list some research questions that may inspire the design of new representations.

\begin{itemize}
    \item What is the potential of Demand-layered MLPTNs to advance our understanding of travel behaviour?
    \item How can multilayer networks be used to represent PT systems at varying levels of abstraction to address diverse planning problems?
    \item What properties of the PT system determine the couplings between layers? How does this translate into setting the weights of inter-layer edges?
    \item How can we use heterogeneous multilayer networks to explore the interactions between different sets of entities in PTNs (stations, routes, passengers, vehicles, ...)?
    \item In what meaningful ways can monolayer networks be re-imagined as multilayer networks and what advantages for understanding and planning of PTNs does it offer?
    \item In what ways can MLPTN modelling be meaningfully inspired by advances in other application domains?
    \item What is the potential of multilayer representations to integrate public transport with other sustainable travel modes (walking, cycling, ride-pooling, ...)?
    \item How can community-layered representations guide the development of X-minute cities?
    \item What is the potential of temporal networks to model and plan dynamic PT systems?
\end{itemize}

\subsection{Public transport planning problems}

Almost all of the considered literature addresses, albeit implicitly, aspects relevant for the long-term strategic planning of PT systems. In addition, the majority of past work is concerned with resilience or basic network statistics. We call for additional research in underrepresented areas across all stages of PT planning: strategic, tactical and operational \cite{ibarra2015planning}.

Strategic PT planning is concerned with Transit network design which is closely intertwined with solving the transit network assignment problem \cite{cats202550}. MLPTN offers several promising avenues for research in this regard:

\begin{itemize}
    \item What is the potential of multilayer networks for accessibility-based planning of public transport systems?
    \item How can multilayer networks be used to understand accessibility at different scales and their integration to offer a unified view of accessibility?
    \item How can we use heterogeneous networks to incorporate the patterns of passenger behaviour in the planning?
\end{itemize}

\color{black}

Tactical planning problems address the need to allocate resources effectively and efficiently which includes topics pertaining to service, vehicle and crew scheduling. Related directions for future research include:

\begin{itemize}
    \item What is the potential of temporal networks to address tactical planning problems?
    \item What heterogeneous network representations can be used to solve resource allocation problems?
    \item How can constraints in optimisation problems be modelled through the use of MLNs?
\end{itemize}

Finally, real-time management involves operational planning and control problems for which we identify the following potential research questions in relation to the potential use of MLPTN:

\begin{itemize}
    \item What functional interactions between nodes that relate to disruptions can be introduced as layers?  Here, functional interactions may pertain to delay propagation.
    \item How can multilayer networks advance the understanding and facilitate the development of control strategies?
    \item What is the potential of multilayer networks to address multi-objective optimisation problems?
    \item What MLN metrics can be used in objective functions of control optimization problems?
\end{itemize}

As resilience analysis represents a significant part of MLPTN studies, we address the specific challenges and opportunities related to it in a dedicated subsection below.

\subsection{Resilience of MLPTNs}
\label{sec:open-resilience}
Resilience studies of MLPTNs differ substantially in how they represent system structure, operations, disruption/recovery duration, and passenger behaviour.
% We analyze these differences and propose possible questions for future research.
An important source of variation concerns modelling granularity.
MLPTN resilience studies vary widely in the level of detail used to represent the transport system. Some works adopt highly detailed infrastructure-based models that include elements such as track layouts, switches, signalling, or catenary systems (e.g.,~\cite{ding_assessing_2025}). Others favour abstract representations in which the network is simplified to stops and inter-stop connections (e.g.,~\cite{gao_understanding_2025}). 
While detailed models allow for greater realism in specific case studies, they often sacrifice generality and limit applicability to other systems, motivating several lines for future research:
%Conversely, more abstract models enhance scalability and cross-system comparability but may overlook operational constraints crucial for realistic disruption scenarios.
\begin{itemize}
    \item What level of infrastructure detail is needed to adequately model different types of disruptions and how can this be encoded into the multilayer structure?
    \item What granularity level allows for greater comparability across networks?
    \item What metrics can be used to quantitatively assess the suitable level of detail?
\end{itemize}

A frequently overlooked aspect in robustness analysis concerns the operational implications of removing a node or link. For example, if a disruption splits a rail or metro line into two disconnected segments, it is not always clear how vehicle operations should be updated: Can vehicles turn back at arbitrary points, or only at designated turn-back facilities? Do service frequencies need adjustment, and how should dwell or turnaround times be handled? Such operational constraints can substantially influence the system's response to disruptions, yet most studies abstract away these details (e.g.,~\cite{palk_robustness_2026,yang_statistic_2014}), potentially overestimating the network's adaptability. This motivates the following research questions:

\begin{itemize}
     \item Which operational constraints have the greatest impact on robustness outcomes?
     \item How can such constraints be most effectively integrated into robustness analysis frameworks?
     \item What is the potential of multilayer network representations to model the different constraints in a single representation?
\end{itemize}

Another source of heterogeneity in the literature lies in how time is represented during simulations. Most works treat disruption and recovery as instantaneous state changes, with few assigning explicit durations to failures and recovery processes (e.g.,~\cite{ju_multilayer_2022,zhong_restoration_2024}). The latter allows modelling the dynamic evolution of system performance over time, including transient phases of degradation and partial restoration. However, time-explicit models are more complex and data-intensive. 
Accordingly, the following research questions emerge:

%The choice between static and dynamic representations reflects a trade-off between simplicity and the ability to capture realistic temporal patterns of disruption propagation and recovery.
\begin{itemize}
    \item What is the potential of multilayer networks to explicitly model functional or causal relationships in failure propagation?
        \item How sensitive are resilience assessments to assumptions about failure and recovery durations?
    \item How can insights from temporal or time-varying network analysis be effectively incorporated into MLPTN resilience analysis?
\end{itemize}

Most (cascading) failure models assume that disrupted passengers reroute their trips based primarily on spatial proximity, i.e., choosing the nearest available station. However, in reality, passenger decisions are influenced by a broader set of factors, including service frequency, reliability, transfer convenience, and connectivity to major destinations. %For instance, two stations located at the same walking distance from a disrupted node may attract very different passenger flows depending on their level of service or their integration within the wider network. 
Some surveyed works include choice modelling approaches to model passenger behaviour (e.g.,~\cite{hu_robustness_2025,yang_temporal_2021}).
Incorporating and enhancing these models is relevant since neglecting these behavioural dimensions can lead to oversimplified propagation patterns and inaccurate estimates of cascading congestion effects. %Incorporating more realistic, utility-based passenger choice models remains a key challenge for capturing how disruptions truly spread across multilayer public transport networks. 
In this regard, we identify the following lines as promising future research:

\begin{itemize}
    \item How can utility-based passenger choice models be best integrated into multilayer network resilience simulations?
        \item How can MLN models be used to encode different rerouting strategies?
    \item To what extent do rerouting assumptions change the outcomes of resilience simulations?
\end{itemize}

These observations highlight both the richness of existing approaches and the need for more systematic frameworks that explicitly balance realism, generality, computational tractability, and operational interpretability.

\section{Concluding remarks}
\label{sec:concl}

By writing this review, we aimed to have offered a glance into the many possibilities that multilayer networks bring to (public) transport research. Inevitably, realising these opportunities requires tackling related challenges. We strived to clarify the most pressing confusions and lay a solid ground for future research. For a glimpse of the possibilities beyond MLNs, we conclude by mentioning that other advanced graph representations, such as hypergraphs or  simplicial complexes, offer complementary frameworks for modelling and analysing (public) transport networks. Embracing these approaches will allow transport researchers to understand and plan (P)TNs while embracing more of their full complexity.

\clearpage

\bibliographystyle{ieeetr} %alpha, apalike, ieeetr
\bibliography{ref.bib}

\end{document}